\documentclass[reprint, 11pt,superscriptaddress,preprintnumbers,nofootinbib,aps]{revtex4}

\usepackage{graphicx}
\usepackage{amsmath,amssymb,amsfonts,amsthm,stmaryrd,mathtools,mathbbol,bm,physics,tensor}
\usepackage{color}
\usepackage{tikz}
\allowdisplaybreaks[1]

\usepackage[bookmarks,linktocpage, colorlinks=true, plainpages = false, citecolor = treegreen,  linkcolor=darkblue, urlcolor = darkblue, filecolor = blue]{hyperref} 

\def\be#1\ee{\begin{align}#1\end{align}}

\def\ba{\begin{eqnarray}}
	\def\ea{\end{eqnarray}}
\def\nn{\nonumber}

\definecolor{darkblue}{rgb}{0., 0.4, 0.8}
\definecolor{cadmiumred}{rgb}{1., 0., 0.22}
\definecolor{treegreen}{rgb}{0., 0.7, 0.3}
\definecolor{orchid}{rgb}{0.7., 0., 0.5}

\newcommand{\note}[1]{{\textcolor{darkblue}{#1}}}

\begin{document}

\title{All $2D$ generalised dilaton theories from $d\geq 4$ gravities}

\author{Johanna Borissova}
\email{j.borissova@imperial.ac.uk}
\affiliation{Abdus Salam Centre for Theoretical Physics, Imperial College London, London SW7 2AZ, UK}

\begin{abstract}
\bigskip
{{\sc Abstract:} We demonstrate that generic two-dimensional Horndeski theories can arise from the reduction of pure gravities in $d \geq 4$ dimensions,
and therefore generic onshell configurations for the two-dimensional metric and scalar field correspond to genuine
$d$-dimensional gravitational vacuum solutions.
	We discuss separately the 
two-dimensional Horndeski theories which can arise from the reduction of $d$-dimensional  generally covariant gravitational actions built only from curvature invariants without covariant derivatives and possessing second-order equations of motion on $2 + (d-2)$ warped-product backgrounds. The discussion is subsequently extended to generic $d$-dimensional gravitational actions with this latter property. We establish a Birkhoff theorem for all gravitational theories whose reduction yields an integrable two-dimensional Horndeski theory, in which case static spherically symmetric solutions satisfy $g_{tt} g_{rr} = -1$ in Schwarzschild gauge whereby the metric function $g_{tt} = -f$ is determined by an algebraic equation. We therefore propose to refer to all such theories as quasi-topological gravities.
These results can be used to show in reverse that any $d$-dimensional  static spherically symmetric and asymptotically flat spacetime  satisfying $g_{tt} g_{rr} = -1$ in Schwarzschild gauge with an invertible dependence of $f$ on the ADM mass  can be reconstructed explicitly
as a vacuum solution to a $d$-dimensional gravitational theory. We discuss examples of regular black holes such as 
the Bardeen spacetime, 
which could not be obtained from polynomial and non-polynomial quasi-topological gravities  involving only curvature invariants without covariant derivatives.
}

\end{abstract}

\maketitle

\tableofcontents

\section{Introduction}

Horndeski theory~\cite{Horndeski:1974wa} is the most general theory for a metric and a scalar field whose field equations are of no more than second order in derivatives. 
In two dimensions one may view the Horndeski action expressed as a generalised Galileon~\cite{Kobayashi:2011nu,Kobayashi:2019hrl} as the most general extension of two-dimensional dilaton theory~\cite{Banks:1990mk}. See the related discussion in~\cite{Grumiller:2021cwg} and also~\cite{Grumiller:2002nm} for a review on generalisations of two-dimensional dilaton theory. 

Generalised two-dimensional dilaton models
may be viewed as effective theories for the  degrees of freedom of $d$-dimensional warped-product spacetimes composed out of a $d-2$ dimensional compact space of constant curvature and a two-dimensional orbit space.
Onshell configurations for the two-dimensional metric and scalar field can then be interpreted as effective $d$-dimensional vacuum solutions. Our particular interest here lies in two-dimensional dilaton models as relevant to the discussion of $d$-dimensional black holes. Geometries in this effective solution space include for instance
$d$-dimensional static regular black holes with spherical or more generally toroidal horizon topology~\cite{Kunstatter:2015vxa}. 
However, such 
spacetimes a priori represent only effective $d$-dimensional vacuum solutions in that their degrees of freedom solve a two-dimensional Horndeski theory rather than a $d$-dimensional generally covariant gravitational theory.~\footnote{
	The variations of the two-dimensional Horndeski action can be used to construct a symmetric second-order and covariantly conserved rank-two tensor as a generalisation of the Einstein tensor~\cite{Lovelock:1971yv}, in terms of which the two-dimensional Horndeski equations of motion can be written as effective generalised Einstein equations on this class of $2+(d-2)$ warped-product backgrounds~\cite{Carballo-Rubio:2025ntd}. Our discussion here shows that this tensor for any choice of two-dimensional metric and scalar field is equivalent to, and can be explicitly reconstructed as, the reduction of the equation of motion tensor of a $d$-dimensional gravitational theory with second-order equations on this class of $2+(d-2)$ warped-product backgrounds.}  \\

Our main goal here is to demonstrate that generic two-dimensional Horndeski theories can arise from the reduction of
generally covariant gravitational theories in $d\geq 4$ dimensions with no additional fields beyond the spacetime metric. Therefore generic onshell configurations of two-dimensional Horndeski theory
in the absence of additional two-dimensional matter degrees of freedom correspond to genuine $d$-dimensional gravitational vacuum solutions. We will focus in particular on the subset of integrable two-dimensional Horndeski theories and corresponding solutions in Schwarzschild gauge, which 
after eliminating a residual time dependence 
 can be 
 ``lifted" as the unique static spherically symmetric solutions satisfying $g_{tt}g_{rr} = -1$  whereby the metric function $g_{tt} =-f(r)$ is determined through an algebraic equation.~\footnote{Here and below we refer to spherically symmetric backgrounds for concreteness, but more generally the reduction can be performed on $2+(d-2)$ warped-product backgrounds with a $d-2$ dimensional transverse compact space other than the sphere.} We will therefore refer to gravitational theories with second-order equations of motion on $2+(d-2)$ warped-product backgrounds satisfying such a Birkhoff theorem as
 {\it quasi-topological gravities} (QTGs). \\

Particular subclasses of two-dimensional Horndeski theories are known to arise from standard notions of QTGs~\cite{Oliva:2010eb,Myers:2010jv,Dehghani:2011vu,Cisterna:2017umf,Bueno:2019ltp,Bueno:2019ycr, Bueno:2022res,Moreno:2023rfl,Bueno:2025qjk} as gravitational theories
built from polynomial functions of curvature invariants without covariant derivatives and possessing second-order equations of motion on spherically symmetric backgrounds~\cite{Bueno:2025qjk}. The two-dimensional Horndeski theories arising from the spherical reduction of such gravities are automatically integrable and in particular the unique static spherically symmetric solutions satisfy $g_{tt}g_{rr} = -1$ whereby $f$ is determined through an algebraic equation involving a single one-variable function of the eigenvalue $\psi(r,f)=(1-f)/r^2$ of the Riemann tensor~\cite{Bueno:2025qjk}. This same function is sufficient to parametrise all two-dimensional Horndeski actions which can arise from the spherical reduction of such theories~\cite{Bueno:2025qjk}. We will henceforth refer to this particular subclass of QTGs as {\it polynomial curvature quasi-topological gravities} (polynomial CQTGs).  
Polynomial CQTGs beyond general relativity exist only in $d\geq 5$ dimensions~\cite{Bueno:2019ltp,Moreno:2023rfl}. \\

CQTGs can be constructed also in $d=4$ dimensions if the action is allowed to depend non-polynomially on curvature invariants~\cite{Bueno:2025zaj,Borissova:2026wmn}. In fact an extended subclass of two-dimensional Horndeski theories can arise from $d\geq 4$ gravitational theories involving non-polynomial functions of curvature invariants without covariant derivatives and possessing second-order equations of motion on spherically symmetric backgrounds.
This subspace of  two-dimensional Horndeski theories  can be 
 subdivided into integrable and non-integrable ones, and we will refer to those
pure-curvature gravities which yield integrable two-dimensional Horndeski theories as {\it non-polynomial curvature quasi-topological gravities} (non-polynomial CQTGs). The metric function $ f$ of static spherically symmetric solutions to these theories is determined through an algebraic equation involving generically two one-variable functions of $\psi(r,f)$. We will establish these statements here for $d \geq 4 $ as an extension of~\cite{Borissova:2026wmn} for $d=4$.\\

Moreover we will demonstrate that generic two-dimensional Horndeski theories can arise from $d \geq 4$ gravitational theories with second-order equations of motion on spherically symmetric backgrounds if the action is allowed in addition to depend on curvature-derivative invariants. This can be understood intuitively as
such invariants can produce separate instances of the variables $f$ and $r$ in $\psi(r,f)$. The gravitational theories of this generic type whose reduction yields an integrable two-dimensional Horndeski theory then provide our definition of all QTGs. We will show that the metric function $f$ of static spherically symmetric solutions in this case is determined through an algebraic equation involving in general two functions of $r$ and $f$ as independent variables.\\

\begin{figure}[htb!]
	\centering
	\includegraphics[width=0.99\textwidth]{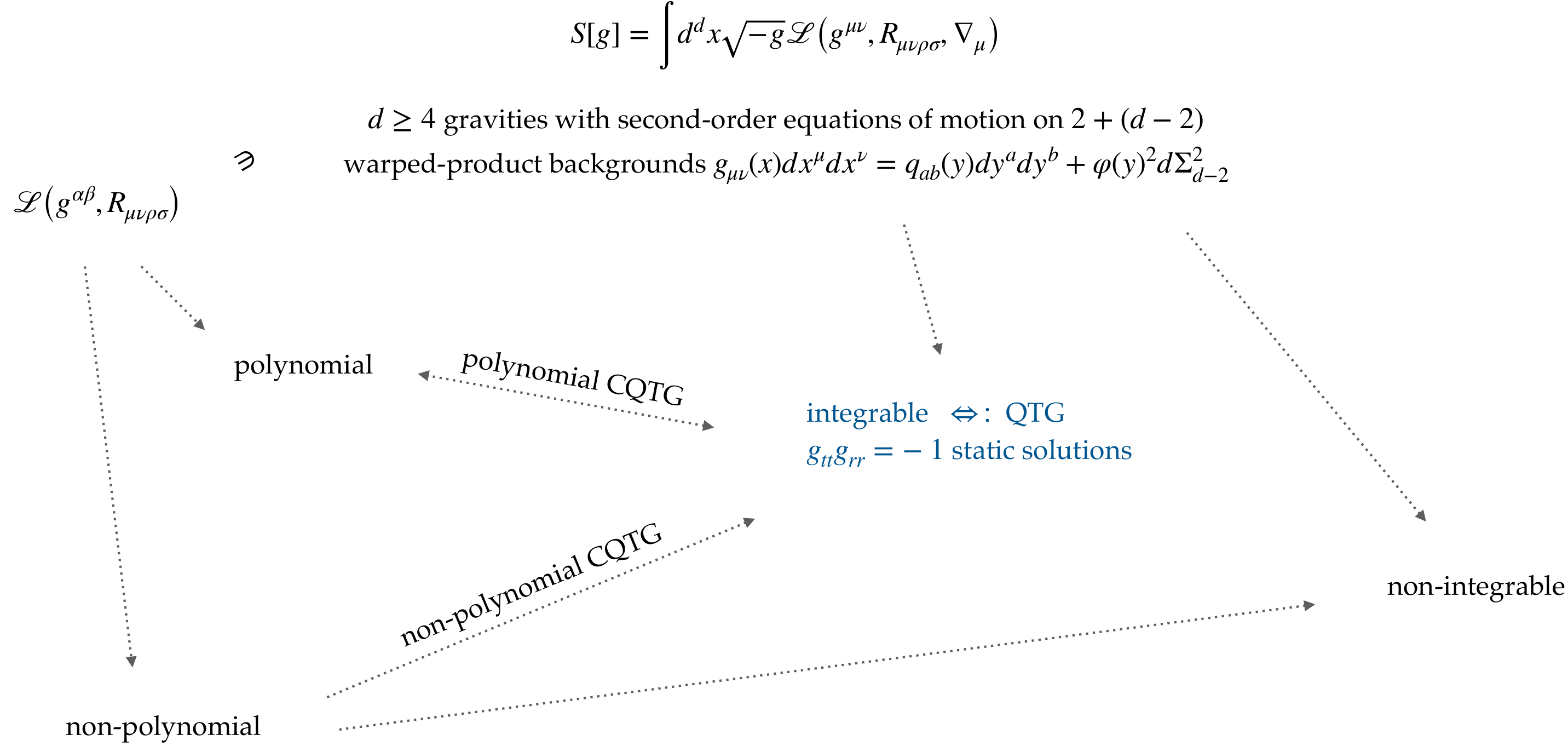}
	\caption{\label{Fig:Diagram} Landscape of $d\geq 4$ gravitational theories  with second-order equations of motion on $2+(d-2)$ warped-product backgrounds.}
\end{figure}

In summary, we can draw the landscape of $d\geq 4$ gravitational theories with second-order equations of motion on spherically symmetric, or more generally on $2+(d-2)$ warped-product backgrounds, shown in Figure~\ref{Fig:Diagram}.\\

These results will allow us moreover to conclude in reverse that any $d$-dimensional static spherically symmetric and asymptotically flat spacetime satisfying $g_{tt}g_{rr} = -1$ in Schwarzschild gauge with an invertible dependence of $f$ on the Arnowitt-Deser-Misner  (ADM) mass
can be reconstructed explicitly as a vacuum solution to a $d$-dimensional generally covariant gravitational theory. To that end, one may relate the ADM mass to the integration constant arising in solutions to the equations of motion of an associated integrable two-dimensional Horndeski theory. Such a reverse construction of solutions to two-dimensional Horndeski theories has been developed in~\cite{Boyanov:2025pes,Carballo-Rubio:2025ntd} and has been applied also in \cite{Borissova:2026dlz,Borissova:2026wmn}. The main statement here then follows from the fact that these two-dimensional Horndeski theories are reduced $d$-dimensional gravities. As a follow up of this central result, in~\cite{Borissova:2026rbi} we discuss the thermodynamics of generic asymptotically flat and anti-de Sitter black holes satisfying $g_{tt}g_{rr}=-1$ in Schwarzschild gauge as solutions to the extended notion of QTGs introduced here, and derive a first law of thermodynamics for any such black hole.\\

The remainder of this article is structured as follows. Section~\ref{Sec:2DHorndeski} provides a review of two-dimensional Horndeski theory. In subsection~\ref{SecSub:ActionEOM} we state the general two-dimensional Horndeski action and covariant equations of motion. In subsection~\ref{SecSub:Birkhoff} we evaluate these in extended Schwarzschild gauge and state the condition for their integrability. The reverse construction of particular classes of solutions to integrable two-dimensional Horndeski theories is summarised in subsection~\ref{SecSub:SolvingIntegrableTheories}. These   discussions are applied exemplarily to a subclass of two-dimensional dilaton actions in subsection~\ref{SecSub:Example2DDilaton}. In section~\ref{Sec:2DHorndeskiFromdDGravity} we discuss how two-dimensional Horndeski actions can be generated from the reduction of $d$-dimensional gravitational actions on $2+(d-2)$ warped-product backgrounds. The reduced action on these backgrounds is displayed in subsection~\ref{SecSub:ReducedAction}. In subsections~\ref{SecSub:CurvatureInvariants} and~\ref{SecSub:Expressing2DHorndeskiCovariant} we discuss curvature and curvature-derivative invariants for these spacetimes and describe a procedure by which generic sets of Horndeski theories arising from the reduction of gravitational actions built from such types of invariants can be generated. As an explicit example, in subsection~\ref{SecSub:ExplicitConstruction4D} we construct  $4$-dimensional Lagrangian densities which are able to produce generic two-dimensional Horndeski theories upon reduction. In section~\ref{Sec:Local} we focus on the subset of Horndeski theories which can arise from the reduction of pure-curvature gravities. Subsection~\ref{SecSub:EOMLocal} derives the general structure of the two-dimensional Horndeski equations of motion which can arise from the reduction of such theories, whereas subsection~\ref{SecSub:IntegrabilityLocal} discusses the integrability condition. This allows us to rederive a Birkhoff theorem for polynomial CQTGs and generalise this theorem to a Birkhoff theorem for non-polynomial CQTGs. In section~\ref{Sec:QuasiLocal} we proceed to discuss the general space of two-dimensional Horndeski theories which can arise from the reduction of curvature-derivative gravities. The equations of motion and integrability condition are discussed in subsections~\ref{SecSub:EOMQuasiLocal} and~\ref{SecSub:IntegrabilityQuasiLocal}. This allows us to derive a  general Birkhoff theorem for any type of QTG. In section~\ref{Sec:SingleFunctionStaticBHs} we illustrate how these results can be applied to reconstruct any static spherically symmetric and asymptotically flat spacetime satisfying $g_{tt}g_{rr}=-1$ and with an invertible dependence of $f$ on the ADM mass as a vacuum solution to a $d$-dimensional gravitational theory. The regular Hayward spacetime is reviewed as a solution to polynomial and non-polynomial CQTGs in subsection~\ref{SecSub:Hayward}. In subsection~\ref{SecSub:Dymnikova} we discuss the Dymnikova spacetime as a solution to non-polynomial CQTGs. Finally, in subsection~\ref{SecSub:Bardeen} we discuss the Bardeen spacetime as a solution which cannot not be obtained from CQTGs but requires the generalisation to  QTGs involving curvature-derivative invariants. We finish with a discussion in section~\ref{Sec:Discussion}.

\section{2D Horndeski theory and equations of motion}\label{Sec:2DHorndeski}

\subsection{Action and covariant equations of motion}\label{SecSub:ActionEOM}

In this subsection we provide a review of two-dimensional Horndeski theory and its covariant equations of motion. Different representations of the action are discussed to emphasise that physical inequivalence of theories must be inferred from the distinctiveness in structure of the equations of motion. This structure will be central to our discussion of two-dimensional Horndeski theories arising from the reduction of $d$-dimensional gravitational theories. \\

Horndeski theory~\cite{Horndeski:1974wa} is the most general theory for a metric and a scalar field with field equations of no more than second order in derivatives. In $1+1$ dimensions the Horndeski action for a metric $q_{ab}(y)$ and a scalar field $\varphi(y)$ can be written as~\cite{Kobayashi:2011nu,Kobayashi:2019hrl}
\ba\label{eq:SHorndeskiLong}
S_{\text{Horndeski}}[q,\varphi] &=& \int \dd[2]{y} \sqrt{-q}\,\bigg[h_2(\varphi,\chi) - h_3(\varphi,\chi) \Box \varphi + h_4(\varphi,\chi)\mathcal{R} \nn\\
&-& 2 \partial_\chi h_4(\varphi,\chi) \qty[\qty(\Box \varphi)^2 - \nabla_a \nabla_b \varphi \nabla^a \nabla^b \varphi]\bigg]\,,
\ea
where $\mathcal{R}$ is the Ricci scalar and
\ba
\chi &=& \nabla_a \varphi \nabla^a \varphi
\ea
the scalar kinetic term. The functions $h_i$ are general functions of $\varphi$ and $\chi$.
The equations of motion are obtained by setting the respective field variations of the action to zero, i.e.,
\ba
\mathcal{E}_{ab} \,\, \vcentcolon =\,\, \frac{1}{\sqrt{-q}} \frac{\delta S_{\text{Horndeski}}}{\var q^{ab}} \,\, = \,\,0 \,, \,\,\, \quad \quad 
\mathcal{E}_{\varphi} \,\, \vcentcolon = \,\, \frac{1}{\sqrt{-q}} \frac{\delta S_{\text{Horndeski}}}{\var \varphi} \,\,= \,\,0\,.\label{eq:EOM}
\ea
These variations satisfy the offshell Bianchi identity
\ba\label{eq:Bianchi}
\nabla^a \mathcal{E}_{ab} + \frac{1}{2} \mathcal{E}_\varphi \nabla_b \varphi &=& 0
\ea
implied by general covariance of the action~\eqref{eq:SHorndeskiLong}. Therefore solving the equation of motion for the metric 
is sufficient to ensure that also the equation of motion for the scalar field is satisfied provided the latter is non-constant. The former can be written as~\cite{Carballo-Rubio:2025ntd}
\ba
\mathcal{E}_{ab} &=& - \frac{1}{2}\big[\alpha + 2\beta \Box \varphi\big]q_{ab}  +
\omega \nabla_a \varphi \nabla_b \varphi  +  \beta \nabla_a \nabla_b \varphi \,\, = \,\, 0\,, \label{eq:Eab}
\ea
where
\ba
\alpha(\varphi,\chi) \,\,= \,\,  h_2 + \chi \partial_\varphi \qty( h_3  - 2 \partial_\varphi h_4)\,, \,\,\, \quad \quad 
 \beta(\varphi,\chi) \,\,=\,\, \chi \partial_\chi \qty(h_3  - 2 \partial_\varphi h_4) - \partial_\varphi h_4 \,,
\label{eq:AlphaBeta}
\ea
and
\ba\label{eq:omega}
\omega(\varphi,\chi) &=& \partial_\chi \alpha - \partial_\varphi \beta \,.
\ea
\\
In two dimensions the last two terms in the Horndeski action~\eqref{eq:SHorndeskiLong} representing the quartic Galileon contribution~\cite{Deffayet:2011gz}, i.e.,
\ba
S_{\text{Galileon}}[q,\varphi] &=& \int \dd[2]{y} \sqrt{-q}\,\qty[h_4(\varphi,\chi)\mathcal{R} - 2 \partial_\chi h_4(\varphi,\chi) \qty[\qty(\Box \varphi)^2 - \nabla_a \nabla_b \varphi \nabla^a \nabla^b \varphi]]\,,
\ea
are redundant and can be recast into a form that resembles the structure of the first two terms by performing partial integrations in~\eqref{eq:SHorndeskiLong}~\cite{Takahashi:2018yzc}. 
This can be anticipated already as  the equations of motion~\eqref{eq:Eab} depend only on two functions $\alpha(\varphi,\chi)$ and $\beta(\varphi,\chi)$ defined in~\eqref{eq:AlphaBeta}. Consider for example the redefinition
\ba
h_2(\varphi,\chi) \quad \to \quad H_2(\varphi,\chi) 
&=&  h_2 + \chi \int \dd{\chi} \qty[\frac{1}{\chi}\partial_\varphi^2 h_4]\,,\label{eq:h2hat}\\
h_3(\varphi,\chi)  \quad \to \quad H_3(\varphi,\chi) 
&=& h_3 -  2 \partial_\varphi h_4 - \int \dd{\chi} \qty[ \frac{1}{\chi} \partial_\varphi h_4]\, ,\label{eq:h3hat}
\ea
which results in 
 \ba\label{eq:AlphaBetaH}
\alpha(\varphi,\chi) \,\, = \,\, H_2 + \chi \partial_\varphi H_3 \,, \,\,\, \quad \quad 
\beta(\varphi,\chi) \,\,=\,\, \chi \partial_\chi H_3\,,
\ea
and therefore effectively eliminates the contribution from the non-minimal coupling $h_4$.
Such redefinitions of the functions $h_i$ amount to performing partial integrations in the action and do not affect the equations of motion. The only physical quantities are the functions $\alpha$ and $\beta$. In particular, 
 the equations of motion derived from the action~\eqref{eq:SHorndeskiLong} where $h_4 $ is set to zero are equivalent to the equations of motion derived from the full action~\eqref{eq:SHorndeskiLong} for appropriately redefined functions $h_2$ and $h_3$. The former representation of the action is identical to the one for kinetic gravity braiding~\cite{Deffayet:2010qz}. Here we purposely choose to display the two-dimensional Horndeski action as in~\eqref{eq:SHorndeskiLong}, as this is the most general functional resulting in second-order equations of motion that can arise from the reduction of $d$-dimensional gravitational actions with second-order equations of motion on $2+(d-2)$ warped-product backgrounds composed from $q_{ab}(y)$ and $\varphi(y)$ prior to performing partial integrations following this reduction. This will allow us later to formulate Birkhoff theorems concerning spherically symmetric solutions to $d$-dimensional QTGs in terms of functions which can be deduced from  the evaluated $d$-dimensional Lagrangian density on such spacetimes, similarly as the Birkhoff theorems for the subclass of polynomial CQTGs derived in~\cite{Bueno:2025qjk}.~\footnote{In general there is an additional term parametrised by a function $h_5(\varphi,\chi)$ in the Horndeski action~\cite{Kobayashi:2011nu,Kobayashi:2019hrl}. This term consists of the Einstein tensor and additional contractions involving covariant derivatives of the scalar field which  vanish in two dimensions.}
 \\

A careful
distinguishment between different representations of the action versus actual inequivalence of corresponding Horndeski theories will be central for what follows. By inequivalent Horndeski theories we will refer to theories $i$ for which the functions $\alpha_i$ as well as the functions $\beta_i$ entering the equations of motion~\eqref{eq:Eab} are distinct up to a possible identical numerical factor arising from the overall normalisation of the action.

\subsection{Equations of motion in extended Schwarzschild gauge and Birkhoff theorem}\label{SecSub:Birkhoff}

In this subsection we state the equations of motion~\eqref{eq:Eab} explicitly in the extended Schwarzschild-type gauge
\ba 
q_{ab}(y)\dd{y}^a \dd{y}^b &=& - n(t,r)^2 f(t,r)\dd{t}^2 + \frac{\dd{r}^2 }{f(t,r)}\,,\label{eq:qabPhiSchwarzschildGauge1}\\
\varphi(y) &=& r \,.\label{eq:qabPhiSchwarzschildGauge2}
\ea
This form of the two-dimensional metric and scalar field will be relevant later for the description of $d$-dimensional spherically symmetric spacetimes, or more generally $2+(d-2)$ warped-product spacetimes with a constant-curvature transverse space. The above ansatz in this case provides a parametrisation for a general dynamical $d$-dimensional line element compatible with these symmetry assumptions and capable of describing a black-hole geometry.\\

The central equations and statements reviewed below have been discussed using  different notation in~\cite{Carballo-Rubio:2025ntd} and in advanced coordinates in~\cite{Borissova:2026dlz,Boyanov:2025pes}. The discussion of the $t$-independent subcase can also be found in~\cite{Borissova:2026wmn}.\\

The scalar kinetic term in the above choice of gauge is given by
\ba
\chi(t,r) &=& f(t,r)\,.
\ea
The independent components of the equations of motion tensor in~\eqref{eq:Eab} are
\ba
\mathcal{E}_{tt} &=& \frac{f n^2}{2}\big[\alpha + \beta \partial_r f] \,\, = \,\,0\,,\label{eq:Ett}\\
\mathcal{E}_{tr} &=& \frac{\beta \partial_t f}{2 f}\,\, = \,\,0\,,\label{eq:Etr}\\
\mathcal{E}_{rr}&=&  -\frac{1}{2   f }\big[\alpha + \beta \partial_r f\big] - \frac{\partial_r n}{n}\beta + \partial_f \alpha - \partial_r \beta 
\,\, =\,\, 0\,,\label{eq:Err}
\ea
where the functional dependence $\alpha(r,f)$ and $\beta(r,f)$ is implied onshell. In the following we will assume that the function $\beta$ is non-zero.~\footnote{This assumption is central for the validity of a Birkhoff theorem for theories satisfying the integrability condition $\partial_\chi \alpha - \partial_\varphi \beta = 0$ offshell. In this case $f=f(r)$ is $t$-independent and the residual $t$-dependence contained in $n = n(t)$ can be gauge-fixed to $n=1$.  If for a theory satisfying the offshell integrability condition, however, $\beta(r,f) =0$ onshell, which fixes $f$, then the first equation implies that also $\alpha(r,f) =0$ onshell, and thus the last equation is automatically satisfied with $n= n(t,r)$ left unconstrained. Vacuum solutions to a $d$-dimensional theory whose reduction yields such a two-dimensional Horndeski theory will be in general non-static spacetimes with $g_{tt}g_{rr}\neq -1$.} Then the second equation requires that $f$ be $t$-independent and thus we continue to write $f =f(r)$. As a result, $\alpha$ and $\beta$ as well as their derivatives do not depend on the coordinate $t$. The remaining two independent equations can then be written as
\ba
\alpha + \beta f' &=& 0\label{eq:Ef}\,,\\
\frac{\partial_r n}{n} &=& \frac{\partial_f \alpha - \partial_r \beta}{\beta}\,.\label{eq:En}
\ea
The second equation can be solved in the form
\ba
n(t,r) &=& \nu(t) \,e^{\int \dd{r}  \frac{\partial_f \alpha - \partial_r \beta}{\beta}}\,,
\ea
where the function $\nu(t)$ can be absorbed by a redefinition of the coordinate $t$. Without loss of generality one may thus set $n = n(r)$ in~\eqref{eq:qabPhiSchwarzschildGauge1} and in particular this function is determined entirely by the solution $f$ of the first equation~\eqref{eq:Ef}. \\

The above discussion shows that generic dynamical configurations of the form~\eqref{eq:qabPhiSchwarzschildGauge1} in the  solution space of the Horndeski equations of motion~\eqref{eq:Eab} are in fact static after gauge-fixing the residual time dependence of $n$~\cite{Carballo-Rubio:2025ntd} --- whereby the notion of staticity becomes meaningful when interpreting two-dimensional Horndeski theory as describing the dynamics of $d$-dimensional spacetimes $g_{\mu\nu}(x)$ composed as a warped product out of the degrees of freedom $q_{ab}(y)$ and $\varphi(y)$. In other words, if the equations of motion of a given two-dimensional Horndeski theory evaluated for this type of ansatz can arise from the reduction of a $d$-dimensional gravitational theory for a spacetime metric, then the therefrom composed vacuum solutions to this gravitational  theory will be necessarily static modulo the above residual time dependence. The possibility of obtaining two-dimensional Horndeski theories from the reduction of $d$-dimensional gravities is the main focus of this paper.\\

Let us emphasise that for configurations of the form~\eqref{eq:qabPhiSchwarzschildGauge1}--\eqref{eq:qabPhiSchwarzschildGauge2} with $n=1$, already imposing $f=f(r)$ according to~\eqref{eq:Etr}, i.e., 
\ba 
q_{ab}(y)\dd{y}^a \dd{y}^b &=& - f(r)\dd{t}^2 + \frac{\dd{r}^2 }{f(r)}\,,\label{eq:qabPhiSchwarzschildGauge1f}\\
\varphi(y) &=& r \,,\label{eq:qabPhiSchwarzschildGauge2f}
\ea
to be a solution to a given two-dimensional Horndeski theory, this theory must satisfy the integrability condition
\ba\label{eq:Integrability}
 \partial_\chi \alpha - \partial_\varphi \beta\,\, = \,\, 0\,.
\ea
Here we use the offshell notation $\alpha(\varphi,\chi)$ and $\beta(\varphi,\chi)$. This statement follows immediately from~\eqref{eq:En}.\\

Vice versa, if a given two-dimensional Horndeski theory satisfies the integrability condition~\eqref{eq:Integrability}, then the unique configurations of the form~\eqref{eq:qabPhiSchwarzschildGauge1}  
in the solution space of this theory are characterised by
\ba
\beta \partial_t f &=& 0\,,\label{eq:Eqft}\\
\beta \partial_r n &=& 0\,,\label{eq:Eqnr}\\
\alpha + \beta \partial_r f &=& 0\label{eq:EqLast}\,,
\ea
whereby these equations hold for generic $\beta$.\\

The previous observations are at the core of uniqueness theorems for spherically symmetric solutions to polynomial CQTGs~\cite{Bueno:2025qjk}. The spherical reduction of such theories yields  two-dimensional Horndeski actions which not only satisfy the integrability condition~\eqref{eq:Integrability}, but also admit a parametrisation of $\alpha$ and $\beta $ in terms of a single one-variable function which determines the function $f$ through an algebraic equation.
We will elaborate on these aspects later and show that for non-polynomial CQTGs the functions $\alpha$ and $\beta$ admit a reparametrisation in terms of two one-variable functions, which results in a generalised algebraic equation for $f$. Finally, we will extend these results to state the most general algebraic equation satisfied by $f$ for generic QTGs.

\subsection{Reverse construction of solutions to integrable theories}\label{SecSub:SolvingIntegrableTheories}

In the following we will demonstrate how configurations characterised by a single function $f$ as in~\eqref{eq:qabPhiSchwarzschildGauge1f} can be reverse-constructed as solutions to a two-dimensional Horndeski theory. This reconstruction will allow us later to generate $d$-dimensional  static spherically symmetric and asymptotically flat spacetimes satisfying $g_{tt}g_{rr}=-1$ with an invertible dependence of $f$ on the ADM mass as vacuum solutions to a $d$-dimensional  generally covariant gravitational theory.\\

 The procedure described below has been developed in~\cite{Boyanov:2025pes,Carballo-Rubio:2025ntd} and has been applied also in~\cite{Borissova:2026dlz,Borissova:2026wmn}.\\

The integrability condition~\eqref{eq:Integrability} which has to be satisfied for the existence of solutions of the form~\eqref{eq:qabPhiSchwarzschildGauge1f} can be solved by means of a characteristic function $\Omega(\varphi,\chi)$ satisfying
\ba\label{eq:IntegrabilityOmega}
\alpha(\varphi,\chi) \,\, = \,\, \partial_\varphi \Omega(\varphi,\chi) \,, \,\,\, \quad \quad \beta(\varphi,\chi) \,\, = \,\, \partial_\chi \Omega(\varphi,\chi)\,.
\ea
With $\varphi = r$ and $\chi = f$ onshell for~\eqref{eq:qabPhiSchwarzschildGauge1f}, the remaining equation of motion~\eqref{eq:EqLast} can then be written as
\ba\label{eq:OmegaEOM}
\derivative{}{r}\qty[\Omega(r,f)] \,\,=\,\, \partial_r \Omega(r,f) + \partial_f \Omega(r,f) f' \,\,= \,\, \alpha(r,f) + \beta(r,f) f' \,\,=\,\,  0\,,
\ea
and can be solved by requiring the function $\Omega(\varphi,\chi)$ to evaluate to an integration constant onshell, i.e.,
\ba\label{eq:OmegaOnshell}
\Omega(r,f) &=& 2 M\,.
\ea
The integration constant $M$ will be related later to the ADM mass of $d$-dimensional static spherically symmetric and asymptotically flat spacetimes.
Notice that for any given theory-dependent choice of functions $\alpha(\varphi,\chi)$ and $\beta(\varphi,\chi)$ satisfying the integrability condition~\eqref{eq:Integrability}, the above is an algebraic equation determining  $f$.\\

For later reference, let us consider concretely how the characteristic function $\Omega(\varphi,\chi)$ defined through~\eqref{eq:IntegrabilityOmega} 
can be expressed in terms of the functions $h_i(\varphi,\chi)$ entering the Horndeski action~\eqref{eq:SHorndeskiLong}. To that end, we observe that the first equation in~\eqref{eq:IntegrabilityOmega} implies
\ba
\Omega(\varphi,\chi) &=& \int \dd{\varphi} \alpha(\varphi,\chi) + F(\chi)
\ea
for an apriori generic function $F$ of $\chi$. Making use of the second equation in~\eqref{eq:IntegrabilityOmega} and invoking~\eqref{eq:Integrability}, it follows however that $F' = 0$, and thus we may set $F =0$ without loss of generality. Any other constant choice of $F$ can be absorbed by a redefinition of the integration constant $2M$. Hence we can express the characteristic function as
\ba\label{eq:Omegahi}
\Omega(\varphi,\chi) &=& \int \dd{\varphi} \alpha(\varphi,\chi) \,\, = \,\, \int \dd{\varphi} \qty[h_2 + \chi \partial_\varphi \qty(h_3 - 2 \partial_\varphi h_4)]\,\,=\,\,
 \chi \qty(h_3 - 2 \partial_\varphi h_4) + \int \dd{\varphi} h_2\,.\quad \quad 
 \ea
Analogously we can conclude
 \ba\label{eq:Omegahi2}
 \Omega(\varphi,\chi) &=& \int \dd{\chi} \beta(\varphi,\chi) \,\,=\,\, \int \dd{\chi} \qty[\chi \partial_\chi \qty(h_3 - 2 \partial_\varphi h_4) - \partial_\varphi h_4] \nn\\
 &=& \chi  \qty(h_3 - 2 \partial_\varphi h_4) - \int \dd{\chi} \qty[h_3 -  \partial_\varphi h_4]\,.
 \ea
Both expressions are equivalent up to a constant, provided the functions $h_i$ satisfy
\ba\label{eq:Integrabilityh}
\partial_\chi h_2 + \partial_\varphi h_3 - \partial_\varphi^2 h_4 &=& 0\,.
\ea
This \note{is} just the integrability condition~\eqref{eq:Integrability} expressed in terms of the functions $h_i$ using the definitions of $\alpha$ and $\beta$ in~\eqref{eq:AlphaBeta}. 
\\

Finally, let us reverse the logic and consider a given a function $f(r)$ involving a constant parameter $M$. Assume the existence of an inverse $M=M(r,f)$. Therefrom, according to~\eqref{eq:OmegaOnshell} one may identify the characteristic function $\Omega(\varphi,\chi) = 2 M(\varphi,\chi)$ offshell and compute the generating functions $\alpha(\varphi,\chi)$ and $\beta(\varphi,\chi)$ from~\eqref{eq:IntegrabilityOmega} which describe this configuration as a solution to the equations of motion~\eqref{eq:Eab} of an integrable two-dimensional Horndeski theory, i.e., for the ansatz~\eqref{eq:qabPhiSchwarzschildGauge1f} and~\eqref{eq:qabPhiSchwarzschildGauge2f}, concretely, equation~\eqref{eq:EqLast}. An explicit realisation of the functions $h_i(\varphi,\chi)$ in the corresponding Horndeski action~\eqref{eq:SHorndeskiLong} is provided by~\cite{Borissova:2026dlz}
\ba\label{eq:hiReverse}
h_2(\varphi,\chi) \,\, = \,\, \alpha(\varphi,\chi)\,,\,\,\,\quad \quad h_3(\varphi,\chi) \,\,=\,\, - 2 \beta(\varphi,\chi) \,,\,\,\,\quad \quad h_4(\varphi,\chi) \,\,=\,\, - \int \dd{\varphi}\beta(\varphi,\chi)\,.
\ea
For this specific  realisation of the functions $h_i$, the first term in brackets in~\eqref{eq:Omegahi} vanishes and therefore in this case the characteristic function can be expressed in terms of $h_2$ as
\ba\label{eq:OmegaReconstruct}
\Omega(\varphi,\chi) &=& \int\dd{\varphi} h_2\,.
\ea

\subsection{Example: 2D dilaton theories}\label{SecSub:Example2DDilaton}

As an examplary application of the previous discussion, let us consider a generalised action for  two-dimensional dilaton (D) theory~\cite{Banks:1990mk}~\footnote{It is well-known that 
	a kinetic term of the form $U(\varphi)\chi$ in the dilaton action~\eqref{eq:SDilaton} can be eliminated by a reparametrisation of the field $\varphi$ together with a $\varphi$-dependent conformal transformation of the metric~\cite{Banks:1990mk,Louis-Martinez:1993bge}. 
Here we do not consider such field redefinitions.
Two theories related by a field redefinition do not need to be physically equivalent and
the physical interpretation of a solution should be performed in the original field variables.} 
\ba\label{eq:SDilaton}
S_{\text{D}}[q,\varphi] &=& \int \dd[2]{y} \sqrt{-q} \,\qty[D(\varphi) \mathcal{R} - \mathcal{V}(\varphi,\chi) ]\,,
\ea
where $D$ denotes a non-minimal coupling of the scalar field to the two-dimensional metric and $\mathcal{V}$ is a generic function of $\varphi$ and $\chi$.
A review on generalised two-dimensional dilaton theories can be found in~\cite{Grumiller:2002nm}.~\footnote{See also~\cite{Nojiri:2000ja} for a review on the quantum properties of these models. We consider two-dimensional dilaton theories as purely classical theories in this work.}
Identifying the functions $h_i$ by comparison with~\eqref{eq:SHorndeskiLong} and computing $\alpha$ and $\beta$ according to~\eqref{eq:AlphaBeta} leads to
\ba
\alpha_{\text{D}}(\varphi,\chi) &=& 
-\mathcal{V}- 2 \chi D''\,,\label{eq:AlphaDilaton}\\
\beta_{\text{D}}(\varphi,\chi) &=&- D'\label{eq:BetaDilaton}\,,
\ea
where a prime denotes the derivative with respect to the argument of a function. Notice that the function $\beta_\text{D}$ for the these particular dilaton models depends only on $\varphi$. For these theories it holds
\ba\label{eq:AlphaBetaIntegrabilityDilaton}
\partial_\chi \alpha_{\text{D}} - \partial_\varphi \beta_{\text{D}}  &=&- \partial_\chi \mathcal{V} - D''\,.
\ea
Thus the integrability condition~\eqref{eq:Integrability} is satisfied iff $D'' =- \partial_\chi \mathcal{V} $. In particular integrability requires $\mathcal{V}$ to depend at most linearly on $\chi$.\\

A prominent example of a dilaton theory satisfying the integrability condition is Jackiw-Teitelboim (JT) gravity~\cite{Teitelboim:1983ux,Jackiw:1984je} with action
\ba\label{eq:SJT}
S_{\text{JT}}[q,\varphi] &=& \int \dd[2]{y}\sqrt{-q}\, \varphi \,\qty[\mathcal{R} + 2 ]\,.
\ea
In this case $\alpha_{\text{JT}} = 2 \varphi$ and $\beta _{\text{JT}}= -1$. \\

Another example satisfying the integrability condition is the Callan–Giddings–Harvey–Stro-minger (CGHS) model~\cite{Callan:1992rs} with action
\ba\label{eq:SCGHS}
S_{\text{CGHS}}[q,\varphi] &=& \int \dd[2]{y}\sqrt{-q} \, e^{-2\varphi}\,\qty[ \mathcal{R} + 4 \chi + 4 \lambda^2 ]\,.
\ea
In this case $\alpha_{\text{CGHS}} = 4 e^{-2\varphi}\qty(\lambda^2 -\chi)$ and $\beta_{\text{CGHS}} = 2 e^{-2\varphi}$.

\section{2D Horndeski theory from $d$-dimensional gravity}\label{Sec:2DHorndeskiFromdDGravity}

Two-dimensional Horndeski theory 
can be viewed as an effective theory for the degrees of freedom of $d$-dimensional warped-product spacetimes
\ba\label{eq:Metric}
g_{\mu\nu}(x) \dd{x^\mu} \dd{x^\nu} &=& q_{ab}(y) \dd{y}^a \dd{y}^b + \varphi(y)^2 \dd{\Sigma_{d-2}^2} \,, \,\,\, \quad \quad \dd{\Sigma_{d-2}^2}  \,\,=\,\,\gamma_{ij}(\theta)\dd{\theta}^i \dd{\theta}^j\,,
\ea
where $\dd{\Sigma_{d-2}^2}$ is the surface element of a 
$d-2$ dimensional compact space of constant sectional curvature $k$ with unit metric denoted by $\gamma_{ij}(\theta)$ and $q_{ab}(y)$ is the metric on a two-dimensional orbit manifold under the isometries of that space. We will here assume that $k$ is non-zero for reasons which will become clear later. Moreover $\varphi(y)$ is a scalar field on this two-dimensional manifold whose value $\varphi =0$ defines the boundary of spacetime. We refer here to spacetimes of the above form as $2+ (d-2)$ warped-product spacetimes and assume $d\geq 4$ throughout.\\

Our goal is to demonstrate that generic two-dimensional Horndeski theories can arise from the reduction of a $d$-dimensional generally covariant action for a spacetime metric
\ba\label{eq:S}
S[g] &=& \int \dd[d]{x} \sqrt{-g}\, \mathcal{L} \,, \,\,\, \quad \quad 
\mathcal{L} \,\, = \,\, \mathcal{L}\big(g^{\mu\nu},R_{\mu\nu\rho\sigma},\nabla_\mu\big)\,.
\ea
The Lagrangian density will be in general a function of curvature invariants without and with covariant derivatives of the metric, and not necessarily analytic. We will implicitly assume its restriction to spacetimes such that the variational principle is well-defined and permits to evaluate the equations of motion for $g$ for an ansatz in terms of $q$ and $\varphi$ of the form~\eqref{eq:Metric}.
Therefrom we can conclude that generic configurations $q$ and $\varphi$ in the solution space of the Horndeski equations of motion~\eqref{eq:Eab} can be associated with spacetimes $g$ composed as in~\eqref{eq:Metric}, which are actual vacuum solutions to a generally covariant $d$-dimensional gravitational theory.

\subsection{Reduced action on $2 + (d-2)$ warped-product backgrounds}\label{SecSub:ReducedAction}

The reduced action~\eqref{eq:S} on warped-product backgrounds~\eqref{eq:Metric} is given by
\ba\label{eq:SReduced}
S[q,\varphi] &=& \mathcal{A}^k_{d-2} \int \dd[2]{y}\sqrt{-q} \,\varphi^{d-2}\,\mathcal{L}(q,\varphi)\,, \,\,\, \quad \quad 
\mathcal{L}(q,\varphi)\,\,=\,\,\eval{\mathcal{L}}_{\eqref{eq:Metric}}\,,
\ea
where $\mathcal{A}_{d-2}^k$ is the constant surface area of the $d-2$ dimensional compact product space. We will drop this overall normalisation factor from now on. The equations of motion for $q$ and $\varphi$ derived from the reduced action~\eqref{eq:SReduced} are equivalent to the equations of motion for $g$ derived from the covariant action~\eqref{eq:SReduced} and evaluated on~\eqref{eq:Metric}. This equivalence of reduced field equations to field equations of the reduced Lagrangian is the principle of symmetric criticality, which applies when the symmetry group is compact~\cite{Palais:1979rca,Fels:2001rv,Deser:2003up,Torre:2010xa,Frausto:2024egp}.\\

To make the translation later between reduced $d$-dimensional gravitational actions and two-dimensional Horndeski actions~\eqref{eq:SHorndeskiLong}, it will be convenient to 
express the latter
in the form
\ba\label{eq:SHorndeskiRewritten}
S_{\text{Horndeski}}[q,\varphi] &=&  
 \int \dd[2]{y} \sqrt{-q} \,\varphi^{d-2}\,\Bigg[{\mathbb{h}}_2(\varphi,\chi) - {\mathbb{h}}_3(\varphi,\chi) \frac{\Box \varphi}{\varphi} + \mathbb{h}_4(\varphi,\chi)\mathcal{R} \nn\\
&-& 2 \partial_\chi \qty[\varphi^2 \mathbb{h}_4(\varphi,\chi) ]\qty[\qty(\frac{\Box \varphi}{\varphi})^2 - \frac{\nabla_a \nabla_b  \varphi}{\varphi} \frac{ \nabla^a \nabla^b \varphi}{\varphi}]\Bigg]\nn\\
&\equiv & \int \dd[2]{y} \sqrt{-q}\, \mathcal{L}_{\text{Horndeski}}(q,\varphi)\,,
\ea
where the functions ${\mathbb{h}}_i(\varphi,\chi)$ are defined as
\ba\label{eq:hhiDef}
{\mathbb{h}}_2(\varphi,\chi) \,\, =\,\, \varphi^{2-d} \,h_2(\varphi,\chi) \,, \,\,\, \quad \quad 
{\mathbb{h}}_3(\varphi,\chi) \,\,=\,\,  \varphi^{3-d}\,h_3(\varphi,\chi) \,,\,\,\, \quad \quad {\mathbb{h}}_4(\varphi,\chi) \,\, =\,\, \varphi^{2-d} \,h_4(\varphi,\chi) \,.\quad 
\ea
Note that with the above definition $\eval{\mathcal{L}}_{\eqref{eq:Metric}} = \varphi^{2-d}\mathcal{L}_\text{Horndeski}$.

\subsection{Invariants for $2+ (d-2)$ warped-product spacetimes}\label{SecSub:CurvatureInvariants}

In this subsection we discuss properties of curvature invariants for the warped-product spacetimes~\eqref{eq:Metric}. Latin indices $a,b,...$ are used to refer to the two-dimensional subspace, whereas $i,j,...$ refer to the $d-2$ dimensional compact transverse subspace.\\

We begin by stating 
the components of the Riemann tensor,
\ba
\eval{R\indices{_a_b_c_d}}_{\eqref{eq:Metric}} \,\,=\,\,  \mathcal{R}\indices{_a_b_c_d}\,,\,\,\,\quad \quad 
\eval{R\indices{_a_i_b_j}}_{\eqref{eq:Metric}} \,\,=\,\, - \varphi \nabla_a\nabla_b \varphi \,\gamma_{ij} \,,\,\,\,\quad \quad 
\eval{R\indices{_i_j_k_l}}_{\eqref{eq:Metric}} \,\,=\,\, 2 \varphi^2\qty(k- \chi ) \gamma_{i[k}\gamma_{l]j}
\,.\label{eq:RiemannComp}\,\,\,\,\,
\ea
The Ricci tensor has components
\ba\label{eq:RicciTensor}
\eval{R_{ab}}_{\eqref{eq:Metric}}  \,\,=\,\, \mathcal{R}_{ab} - (d-2) \frac{\nabla_a\nabla_b \varphi}{\varphi}\,,\,\,\,\quad \quad 
\eval{R_{ij}}_{\eqref{eq:Metric}} \,\,=\,\,  \qty[-\varphi \Box \varphi + (d-3) \qty(k-\chi )]\,\gamma_{ij}\,,
\ea
and the Ricci scalar is given by
\ba\label{eq:R}
\eval{R}_{\eqref{eq:Metric}} &=& \mathcal{R} - 2 (d-2)  \frac{\Box \varphi}{\varphi} +  (d-3)(d-2) \frac{k-\chi}{\varphi^2}\,.
\ea
Therefrom the components of the Weyl tensor are obtained as
\ba\label{eq:WeylTensor}
\eval{C_{abcd}}_{\eqref{eq:Metric}} &=& \frac{d-3}{d-1} \,\sigma \,q_{a[c} q_{d]b}\,,\,\,\,\quad \quad 
\eval{C_{aibj}}_{\eqref{eq:Metric}} \,\,=\,\, - \frac{d-3}{2(d-2)(d-1)} \,\sigma\, \varphi^2 q_{ab} \gamma_{ij}\,,\\
\eval{C_{ijkl}}_{\eqref{eq:Metric}}  &=& \frac{2 }{(d-2)(d-1)} \,\sigma\, \varphi^4 \gamma_{i[k}\gamma_{l]j}\,,\label{eq:WeylTensor2}
\ea
where
\ba\label{eq:Omega}
\sigma &=& \mathcal{R} + 2 \frac{\Box \varphi}{\varphi} + 2 \frac{k-\chi}{\varphi^2}\,.
\ea
The Weyl tensor vanishes in dimensions lower than four and will be used later for the construction of $d$-dimensional covariant densities, which is the reason for our assumption $d\geq 4$. In the previous expressions $\mathcal{R}_{abcd}$, $\mathcal{R}_{ab}$ and $\mathcal{R}$ denote the Riemann tensor, Ricci tensor and Ricci scalar of the two-dimensional metric $q_{ab}$.~\footnote{Remember that in two dimensions the Riemann tensor consists only of the trace component, i.e,
	\ba
	\mathcal{R}_{abcd} &=& \mathcal{R} \, q_{a [c} q_{d] b}\,.
	\ea
} \\

For future reference we also state the inverse spacetime metric used to raise indices. It has the block-diagonal form
\ba\label{eq:MetricInverse}
g^{\mu\nu} &=& \text{diag}\qty{g^{ab},  g^{ij}} \,\, = \,\, \text{diag}\qty{q^{ab}, \frac{1}{\varphi^2} \gamma^{ij}}\,.
\ea
\\
From now on we will use the following terminology. We will refer to curvature invariants constructed from the Riemann tensor without covariant derivatives as {\it curvature invariants}, and will refer to curvature invariants involving covariant derivatives explicitly as {\it curvature-derivative invariants}.

\subsubsection{Curvature invariants}\label{SecSubSub:LocalInvariants}

From the expressions for the components of the Riemann tensor~\eqref{eq:RiemannComp} 
and the inverse metric~\eqref{eq:MetricInverse}
it follows that all curvature invariants for the warped-product spacetimes~\eqref{eq:Metric} are functions of the four scalars
\ba\label{eq:BuildingBlocks}
\qty{\mathcal{R}\,,\, \frac{\Box \varphi}{\varphi}\,,\, \frac{ \nabla_a \nabla_b \varphi}{\varphi} \frac{ \nabla^a \nabla^b \varphi}{\varphi} \,,\,\frac{k-\chi}{\varphi^2}} \,.
\ea
In particular such invariants depend on $\chi$ only through the combination
\ba\label{eq:Psi}
\psi(\varphi,\chi)\,\, \coloneqq\,\, \frac{k - \chi}{\varphi^2}\,.
\ea
This quantity can be interpreted physically as the Misner-Sharp mass density~\cite{Misner:1964je} for spacetimes~\eqref{eq:Metric} in general relativity.\\

{\it Remark:} The fact that curvature invariants cannot produce instances of $\chi$ alone can be understood physically as follows. Consider a static spacetime with line element 
\ba\label{eq:MetricStatic}
g_{\mu\nu}(x) \dd{x}^\mu \dd{x}^\nu &=& - f(r)\dd{t}^2 + \frac{\dd{r}^2}{f(r)} + R(r)^2 \dd{\Sigma}_{d-2}^2\,.
\ea
The kinetic term 
of the scalar field $\varphi(y) = R(r)$ in this case is given by
\ba
\chi(r) &=& f(r) R'(r)^2
\ea
and vanishes on event horizons and wormhole throats of~\eqref{eq:MetricStatic}. The existence of a curvature invariant proportional to $\chi$ would not be compatible with the local equivalence principle.

\subsubsection{Curvature-derivative invariants}\label{SecSubSub:QuasiLocalInvariants}

Curvature-derivative invariants of the warped-product spacetimes~\eqref{eq:Metric} will be in general functions of additional scalar combinations beyond the four in~\eqref{eq:BuildingBlocks}. One such combination is 
\ba\label{eq:Phi}
\phi(\varphi,\chi) \,\, \coloneqq\,\, \frac{ \chi}{\varphi^2}\,.
\ea
This can be seen for instance by computing the square of the covariant derivative of the Weyl tensor in~\eqref{eq:WeylTensor}--\eqref{eq:WeylTensor2}.
Alternatively one may see this by forming scalar contractions  involving covariant derivatives out of the four combinations in~\eqref{eq:BuildingBlocks} such as $\nabla_\mu \psi \nabla^\mu \psi$.
Later in subsection~\ref{SecSub:ExplicitConstruction4D} we will use that there exist curvature-derivative invariants proportional to $\phi$.\\

{\it Remark:} Notice that $\phi$ vanishes on event horizons and wormhole throats of~\eqref{eq:MetricStatic}. The  existence of a coordinate invariant which vanishes on event horizons of static and more generally stationary spacetimes is only possible because an event horizon in a stationary spacetime is also an apparent horizon. The latter notion refers to a quasi-local surface characterised solely by the behavior of the expansions of ingoing and outgoing null geodesics~\cite{Ashtekar:2004cn,Booth:2005qc,Gourgoulhon:2008pu}. In dynamical spacetimes the notions of an event and apparent horizon are generally distinct, and only apparent horizons, or more generally geometric surfaces~\cite{McNutt:2021esy}, can be detected by means of particular combinations of curvature-derivative invariants~\cite{Karlhede:1982fj,Tammelo1997,Gass:1998nd,Mukherjee:2002ba,Lake:2003qe,Saa:2007ub,Gomez-Lobo:2012ibv,Moffat:2014aqa,Visser:2014zqa,Abdelqader:2014vaa,Page:2015aia,McNutt:2017gjg,McNutt:2017paq,Coley:2017vxb,McNutt:2021esy}.~\footnote{This observation has been used in~\cite{Borissova:2024hkc} based on a dynamical suppression mechanism in Lorentzian path integrals described in~\cite{Borissova:2020knn,Borissova:2023kzq} to emphasise that the no-global-symmetry conjecture in quantum gravity~\cite{Banks:1988yz,Giddings:1987cg,Lee:1988ge,Abbott:1989jw,Coleman:1989zu,Kamionkowski:1992mf,Holman:1992us,Kallosh:1995hi,Banks:2010zn} may not  hold universally across generic effective field theories, and in particular those featuring non-localities. See also~\cite{Eichhorn:2024rkc,Basile:2025zjc} for related discussions.}

\subsection{Expressing 2D Horndeski densities as reduced $d$-dimensional covariant densities}\label{SecSub:Expressing2DHorndeskiCovariant}

\subsubsection{Curvature Lagrangian densities}\label{SecSubSub:LocalNonPolynomialDensities}

In the following we will describe an abstract procedure that allows us to generate the generic sets of two-dimensional Horndeski theories which can arise from the reduction of a $d$-dimensional Lagrangian density built only from curvature invariants.
Such an approach has been considered for $d=4$ in~\cite{Borissova:2026wmn}. Here we consider its generalisation applicable for $d\geq 4$.\\

 From the discussion in subsection~\ref{SecSubSub:LocalInvariants} and the Riemann scalars~\eqref{eq:BuildingBlocks}, we see that combinations of the first three scalars can produce the tensorial structures multiplying $-\mathbb{h}_3$, $\mathbb{h}_4$ and $-2 \partial_\chi\qty[\varphi^2  \mathbb{h}_4 ]$ in square brackets of the action functionals~\eqref{eq:SHorndeskiRewritten}. By taking any four independent curvature invariants for the spacetimes~\eqref{eq:Metric} and solving for the four scalars~\eqref{eq:BuildingBlocks} one may therefore generate $d$-dimensional curvature Lagrangian densities whose evaluation on~\eqref{eq:Metric} produces a Horndeski action~\eqref{eq:SHorndeskiRewritten} with functions $\mathbb{h}_i$ constrained to depend on $\varphi$ and $\chi$ in the form
\ba\label{eq:Constrainth}
\mathbb{h}_i(\varphi,\chi) &=& \mathbb{h}_i(\psi)\,.
\ea
This constraint must be understood prior to performing partial integrations following the reduction. Let us denote by $\mathcal{I}_i$ below four $d$-dimensional covariant densities whose evaluation on~\eqref{eq:Metric} yields the scalars~\eqref{eq:BuildingBlocks},
\ba
\eval{\mathcal{I}_{\mathcal{R}}}_{\eqref{eq:Metric}} &=& \mathcal{R}\,,\label{eq:IR}\\
\eval{\mathcal{I}_{\frac{\Box \varphi}{\varphi}}}_{\eqref{eq:Metric}} &=& \frac{\Box \varphi}{\varphi}\,,\label{eq:IBoxVarphiVarphi}\\
\eval{\mathcal{I}_{ \frac{ \nabla_a \nabla_b \varphi}{\varphi} \frac{ \nabla^a \nabla^b \varphi}{\varphi}}}}_{\eqref{eq:Metric} &=& \frac{ \nabla_a \nabla_b \varphi}{\varphi} \frac{ \nabla^a \nabla^b \varphi}{\varphi}\,,\label{eq:IDDVarphiDDVarphi}\\
\eval{\mathcal{I}_{\psi}}}_{\eqref{eq:Metric} &=& \frac{k-\chi}{\varphi^2}\label{eq:IPsi}\,.
\ea
Then a $d$-dimensional gravitational action which produces the subclass of Horndeski actions~\eqref{eq:SHorndeskiRewritten} satisfying the constraint~\eqref{eq:Constrainth} is 
\ba
S[g] &=& \int \dd[d]{x} \sqrt{-g}\,\qty[\mathbb{h}_2\qty(\mathcal{I}_{\psi})- \mathbb{h}_{3} \qty(\mathcal{I}_\psi) \mathcal{I}_{\frac{\Box \varphi}{\varphi}} + \mathbb{h}_4\qty(\mathcal{I}_\psi)\mathcal{I}_{\mathcal{R}} + 2 \mathbb{h}_4'\qty(\mathcal{I}_\psi) \qty[\mathcal{I}_{\frac{\Box \varphi}{\varphi}}^2 - \mathcal{I}_{ \frac{ \nabla_a \nabla_b \varphi}{\varphi} \frac{ \nabla^a \nabla^b \varphi}{\varphi}}]]\,.\quad \,\,\,\label{eq:SLocal}
\ea
The functions $\mathbb{h}_i\qty(\mathcal{I}_{\psi})$ can be chosen arbitrarily to produce any desired  Horndeski actions~\eqref{eq:SHorndeskiRewritten} with $\mathbb{h}_i(\varphi,\chi) = \mathbb{h}_i(\psi)$ upon reduction on~\eqref{eq:Metric}.\\

This is of course a cumbersome way of generating two-dimensional Horndeski theories and it 
requires in general non-polynomial $d$-dimensional covariant densities $\mathcal{I}_i$ which may not be well-defined on arbitrary backgrounds. As emphasised previously, we only consider these non-polynomial theories on backgrounds for which the variational principle is well-defined.
There do not exist polynomial $d$-dimensional curvature Lagrangian densities which upon reduction on~\eqref{eq:Metric} can produce the entire class of two-dimensional Horndeski actions~\eqref{eq:SHorndeskiRewritten} characterised by~\eqref{eq:Constrainth} --- the only two-dimensional Horndeski actions which can arise from the reduction of polynomial curvature gravities are those generated by polynomial CQTGs~\cite{Bueno:2025qjk}.
In this case the functions $\mathbb{h}_i(\psi)$ identified after the reduction can be parametrised in terms of a single analytic function of $\psi$. This parametrisation singles out a strictly smaller subset of all Horndeski actions~\eqref{eq:SHorndeskiRewritten} satisfying~\eqref{eq:Constrainth}. It is therefore necessary to go beyond polynomial 
curvature gravities
 in order to exhaust the space of two-dimensional Horndeski theories which can arise from the reduction of pure-curvature gravities. That this is in principle possible is the main statement here.\\

An exemplary realisation of the above construction has been performed for $d=4$ in~\cite{Borissova:2026wmn} and is reviewed in subsection~\ref{SecSub:ExplicitConstruction4D}. Therein we  will further generalise this construction to generate generic two-dimensional Horndeski theories from $4$-dimensional gravities by allowing in addition curvature-derivative invariants in the action. These considerations can be straightforwardly generalised to $d\geq 4$. The concrete realisation of the covariant densities $\mathcal{I}_i$ is however irrelevant for later our purposes. 

\subsubsection{Curvature-derivative Lagrangian densities}\label{SecSubSub:QuasiLocalNonPolynomialDensities}

The procedure described in the previous subsection can be extended to generate generic two-dimensional Horndeski theories from a reduction of $d$-dimensional curvature-derivative Lagrangian densities. To that end, it only remains to see that a arbitrary dependence on $\varphi$ and $\chi$ of the functions
$
\mathbb{h}_i(\varphi,\chi)
$
in the Horndeski action~\eqref{eq:SHorndeskiRewritten} can be achieved.  
It is possible to 
construct curvature-derivative covariant densities which upon evaluation on~\eqref{eq:Metric} produce the scalar $\phi$ defined in~\eqref{eq:Phi}, i.e.,
\ba
\eval{\mathcal{I}_{\phi}}_{\eqref{eq:Metric}} &=& \frac{\chi}{\varphi^2} \label{eq:IPhi}\,.
\ea
By forming the inverse sum of any given realisation of the invariants $\mathcal{I}_\psi$ and $\mathcal{I}_\phi$ one may thus construct an invariant satisfying~\footnote{These considerations assume $k$ to be non-zero, as stated previously. If $k=0$, it would be $\psi = \phi$ and thus $\varphi$ and $\chi$ cannot be separated in this way.} 
\ba
\eval{\mathcal{I}^2_{\varphi}}_{\eqref{eq:Metric}} &\vcentcolon =&  \eval{\frac{k}{\mathcal{I}_\psi + \mathcal{I}_\phi}}_{\eqref{eq:Metric}}\,\, = \,\,  \varphi^2 \label{eq:IVarphiSq}\,.
\ea
This squared invariant multiplied by $\mathcal{I}_\phi$ in turn can be used to construct an invariant satisfying
\ba
\eval{\mathcal{I}_{\chi}}_{\eqref{eq:Metric}} &\vcentcolon =&  \eval{\mathcal{I}_\phi\, \mathcal{I}_{\varphi}^2}_{\eqref{eq:Metric}} \,\, = \,\, \chi \label{eq:IChi}\,.
\ea
Then a $d$-dimensional gravitational action which produces generic Horndeski actions~\eqref{eq:SHorndeskiRewritten} with functions $\mathbb{h}_i(\varphi,\chi)$ upon reduction on~\eqref{eq:Metric} is
\ba
S[g] &=& \int \dd[d]{x} \sqrt{-g}\,\bigg[\mathbb{h}_2\qty(\mathcal{I}_{\varphi},\mathcal{I}_{\chi})- \mathbb{h}_{3} \qty(\mathcal{I}_{\varphi},\mathcal{I}_{\chi}) \mathcal{I}_{\frac{\Box \varphi}{\varphi}} + \mathbb{h}_4\qty(\mathcal{I}_{\varphi},\mathcal{I}_{\chi})\mathcal{I}_{\mathcal{R}} \nn\\
&-& 2\, \mathcal{I}_{\varphi}^2\,\partial_{\mathcal{I}_\chi}\mathbb{h}_4\qty(\mathcal{I}_{\varphi},\mathcal{I}_{\chi}) \qty[\mathcal{I}_{\frac{\Box \varphi}{\varphi}}^2 - \mathcal{I}_{ \frac{ \nabla_a \nabla_b \varphi}{\varphi} \frac{ \nabla^a \nabla^b \varphi}{\varphi}}]\bigg]\,.\quad \quad \label{eq:SQuasiLocal}
\ea
Notice that this particular construction requires specifying the non-zero constant sectional curvature $k$ of the compact transverse space with respect to which the reduction will be performed, as this constant enters the Riemann scalar $\psi$ in~\eqref{eq:Psi} and is needed in order to separate the variables $\varphi^2$ and $\chi$ through~\eqref{eq:IVarphiSq} and~\eqref{eq:IChi}.\\

The covariant actions arising from the construction above even for simple choices of functions $\mathbb{h}_i$ are generally complicated non-polynomial actions of curvature and curvature-derivative invariants. As  emphasised previously, here we are only interested in a proof of principle that generic two-dimensional Horndeski theories can be generated in this way from the reduction of a $d$-dimensional generally covariant gravitational action.

\subsection{Example: Explicit construction in $d=4$}\label{SecSub:ExplicitConstruction4D}

For completeness, in this subsection we provide an explicit example for a  construction of $d$-dimensional gravitational theories which can produce a given two-dimensional Horndeski theory~\eqref{eq:SHorndeskiRewritten} upon reduction on~\eqref{eq:Metric}.
We  will focus here concretely  on $d=4$, but this construction  can be generalised to $d\geq 4$. Thus, we consider $2+2$ warped-product spacetimes
\ba \label{eq:Metric4d}
g_{\mu\nu}(x)\dd{x}^\mu \dd{x}^\nu &=& q_{ab}(y ) \dd{y}^a \dd{y}^b + \varphi(y)^2 \dd{\Sigma}_{2}^2\,.
\ea
Explicit realisations of the scalars $\mathcal{I}_i$ in~\eqref{eq:IR}--\eqref{eq:IPsi} have been constructed in~\cite{Borissova:2026wmn} and are sufficient to generate the subclass of two-dimensional Horndeski theories which can arise from the reduction of curvature Lagrangian densities.~\footnote{The construction in~\cite{Borissova:2026wmn} assumed spherical symmetry, i.e., $k=1$, but its application extends to generic $k$ as the derivation depends only on the scalar quantity $\psi$.} Here we review the relevant definitions and supplement these by providing explicit realisations for the scalars $\mathcal{I}_\phi$, $\mathcal{I}_\varphi$ and $\mathcal{I}_\chi$ in~\eqref{eq:IPhi}--\eqref{eq:IChi} which allow us to generate the general set of two-dimensional Horndeski theories from $4$-dimensional gravitational actions~\eqref{eq:SQuasiLocal}.\\

Let us define the following curvature invariants,
\begin{center}
	\begin{minipage}{0.4\textwidth}
\ba
\mathcal{I}_{R}  &=& R\,,\\
\mathcal{I}_{\hat{R}^2} &=& \hat{R}\indices{_\mu^\nu} \hat{R}\indices{_\nu^\mu}\,,\\
\mathcal{I}_{C^2} &=& C_{\mu\nu\rho\sigma}C^{\mu\nu\rho\sigma}\,,
\ea
\end{minipage}
\hspace{0.4cm}
\begin{minipage}{0.5\textwidth}
\ba
\mathcal{I}_{\hat{R}^3} &=& \hat{R}\indices{_\mu^\nu} \hat{R}\indices{_\nu^\rho} \hat{R}\indices{_\rho^\mu}\,,\\
\mathcal{I}_{C^3}  &=& C\indices{_\mu_\nu ^\rho ^\sigma} C\indices{_\rho_\sigma^\alpha^\beta}C\indices{_\alpha_\beta^\mu^\nu} \,,\\
\mathcal{I}_{R^2 C}  &=& \hat{R}\indices{^\mu^\nu}\hat{R}\indices{^\rho^\sigma} C_{\rho\mu\nu\sigma}\,,
\ea
	\end{minipage}
\hspace{0.1cm}
\end{center}
where $\hat{R}$ denotes the traceless Ricci tensor, i.e.,
\ba
\hat{R}_{\mu\nu} &=& R_{\mu\nu} - \frac{1}{4} Rg_{\mu\nu} \,.
\ea
Then an exemplary choice of invariants $\mathcal{I}_i$ defined through~\eqref{eq:IR}--\eqref{eq:IPsi} is~\cite{Borissova:2026wmn}
\ba\label{eq:IiLocal}
\mathcal{I}_{\mathcal{R}} &=&\frac{1}{ \mathcal{I}_{C^2} \mathcal{I}_{\hat{R}^2C} + 2\mathcal{I}_{\hat{R}^2} \mathcal{I}_{C^3} }\bigg[\frac{1}{6} \mathcal{I}_{R} \mathcal{I}_{C^2} \mathcal{I}_{\hat{R}^2 C}  +  \frac{1}{3} \mathcal{I}_{R}\mathcal{I}_{\hat{R}^2}  \mathcal{I}_{C^3}\nn\\
 &+& 4 \frac{ \mathcal{I}_{\hat{R}^2} \mathcal{I}_{C^3}^2}{\mathcal{I}_{C^2}} 	+ 2 \mathcal{I}_{\hat{R}^3} \mathcal{I}_{C^3} + 2  \mathcal{I}_{C^3} \mathcal{I}_{\hat{R}^2 C}\bigg]\,,\label{eq:IR4d} \\
\mathcal{I}_{\frac{\Box \varphi}{\varphi}} &=&  -\frac{1}{6} \mathcal{I}_R  +\frac{\mathcal{I}_{C^3}}{\mathcal{I}_{C^2}}\,,\\
\mathcal{I}_{ \frac{ \nabla_a \nabla_b \varphi}{\varphi} \frac{ \nabla^a \nabla^b \varphi}{\varphi}} &=& \frac{1}{72}\mathcal{I}_{R}^2 +\frac{1}{6} \mathcal{I}_{\hat{R}^2} - \frac{1}{6}\frac{\mathcal{I}_R \mathcal{I}_{C^3}}{\mathcal{I}_{C^2}} + \frac{1}{2}\frac{\mathcal{I}_{C^3}^2}{\mathcal{I}_{C^2}^2} + \frac{1}{12}\frac{\mathcal{I}_{C^2}\mathcal{I}_{\hat{R}^2 C}}{\mathcal{I}_{C^3}}\,,\\
\mathcal{I}_{\psi} &=& \frac{1}{12}\mathcal{I}_R + \frac{\mathcal{I}_{C^3}}{\mathcal{I}_{C^2}} -  \frac{\mathcal{I}_{\hat{R}^3}\mathcal{I}_{C^3}}{\mathcal{I}_{C^2} \mathcal{I}_{\hat{R}^2 C} + 2 \mathcal{I}_{\hat{R}^2} \mathcal{I}_{C^3}}\,.\label{eq:IPsi4d}
\ea
The action~\eqref{eq:SLocal} built from these invariants with $d=4$ upon reduction on~\eqref{eq:Metric4d} then yields the subclass of two-dimensional Horndeski actions~\eqref{eq:SHorndeskiRewritten} with generic functions $\mathbb{h}_i = \mathbb{h}_i(\psi)$. \\

Now we extend this construction by stating an invariant $\mathcal{I}_\phi$ satisfying the defining relation~\eqref{eq:IPhi} and following the steps described in the previous subsection to arrive at an action~\eqref{eq:SQuasiLocal} whose reduction produces~\eqref{eq:SHorndeskiRewritten}. To that end, consider in addition the following curvature-derivative invariants,
\ba
\mathcal{I}_{\nabla C \cdot \nabla C} &=& \nabla_\alpha C_{\mu\nu\rho\sigma} \nabla^\alpha C^{\mu\nu\rho\sigma}\,,\\
\mathcal{I}_{\nabla C^2 \cdot \nabla C^2}&=& \nabla_\alpha \mathcal{I}_{C^2} \nabla^\alpha \mathcal{I}_{C^2}\,.
\ea
Then an exemplary realisation of the invariant $\mathcal{I}_\phi$ defined through~\eqref{eq:IPhi} is~\footnote{This invariant has been constructed in~\cite{McNutt:2021esy} as a detector for black-hole apparent horizons and wormhole throats in spherical symmetry. Concretely, for a spherically symmetric line element in advanced coordinates of the form
\ba\label{eq:Metric4dAdvanced}
g_{\mu\nu}(x)\dd{x}^\mu \dd{x}^\nu &=& -n(v,r)^2f(v,r) \dd{v}^2 + 2 n(v,r)\dd{v}\dd{r} + R(r)^2 \dd{\Omega^2}\,,
\ea
this invariant evaluates to
\ba
\eval{\mathcal{I}_\phi}_{\eqref{eq:Metric4dAdvanced}} &=& \frac{f R'^2}{R^2}\,,
\ea
and thus vanishes whenever $f=0$ or $R' =0$.
}
\ba
\mathcal{I}_{\phi} &=& \frac{1}{24
	\mathcal{I}_{C^2}^2} \big[4 \,\mathcal{I}_{C^2}\, \mathcal{I}_{\nabla C \cdot \nabla C} - \mathcal{I}_{\nabla C^2 \cdot \nabla C^2}\big]\,.\label{eq:IPhi4d}
\ea
To see this explicitly, one may use~\eqref{eq:WeylTensor}--\eqref{eq:WeylTensor2} for the covariant components of the Weyl tensor and compute its contravariant components using the inverse metric in~\eqref{eq:MetricInverse}. It is then straightforward to see
\ba
\eval{\mathcal{I}_{C^2}}_{\eqref{eq:Metric4d}} &=& \frac{1}{3}\sigma^2\,,\\
\eval{ \mathcal{I}_{\nabla C \cdot \nabla C} }_{\eqref{eq:Metric4d}} &=&  \frac{1}{3}\nabla_e \sigma \nabla^e \sigma + 2 \sigma^2 \frac{\chi}{\varphi^2}\,,\\
\eval{\mathcal{I}_{\nabla C^2 \cdot \nabla C^2}}_{\eqref{eq:Metric4d}} &=&  \frac{4}{9} \sigma^2\nabla_e \sigma \nabla^e \sigma \,,
\ea
where $\sigma$ was defined in~\eqref{eq:Omega}. Therefore, altogether
\ba
\eval{\mathcal{I}_\phi}_{\eqref{eq:Metric4d}} &=& \frac{\chi}{\varphi^2} \,\,=\,\, \phi\,.
\ea
By using the exemplary realisations of the invariants $\mathcal{I}_\psi$ and $\mathcal{I}_\phi$ provided above, one may construct invariants $\mathcal{I}_{\varphi}$ and $\mathcal{I}_\chi$ satisfying the defining relations~\eqref{eq:IVarphiSq} and~\eqref{eq:IChi}. Therefrom we obtain a $4$-dimensional generally covariant action~\eqref{eq:SQuasiLocal} whose reduction on~\eqref{eq:Metric4d} produces a given desired two-dimensional Horndeski action~\eqref{eq:SHorndeskiRewritten}.~\footnote{As an example, the JT action~\eqref{eq:SJT} can then be obtained from the reduction on~\eqref{eq:Metric4d} of a $4$-dimensional generally covariant action
\ba
S_{\text{JT}}\qty[g] &=& \int \dd[4]{x}\sqrt{-g} \frac{1}{\mathcal{I}_\varphi}\qty[ \mathcal{I}_{\mathcal{R}} + 2]\,,
\ea
with $\mathcal{I}_\mathcal{R}$ given in~\eqref{eq:IR4d} and $\mathcal{I}_\varphi = \qty[\frac{k}{\mathcal{I}_\psi +\mathcal{I}_\phi} ]^{\frac{1}{2}}$ with $\mathcal{I}_\psi$ and $\mathcal{I}_\phi$ given in~\eqref{eq:IPsi4d} and~\eqref{eq:IPhi4d}.}\\

Before closing this subsection, let us mention that related observations concerning the possibility of lifting two-dimensional Horndeski theories to $4$-dimensional non-polynomial gravities have been made in~\cite{Colleaux:2017ibe} and have been generalised to $d\geq 4$ in~\cite{Colleaux:2019ckh}.
	The discussions therein are focused on the particular subcase when the reduced  theory features second-order equations of motion with functions $\mathbb{h}_i$ which are polynomial in $\phi$ and $\phi + \psi$. A similar statement applies to the  $4$-dimensional non-polynomial curvature gravities constructed in~\cite{Bueno:2025zaj} for which the functions $\mathbb{h}_i$ in the reduced action are polynomial in $\psi$ at each finite effective order $n$ in the curvature.~\footnote{See concretely equations $(24)$--$(26)$ in~\cite{Bueno:2025zaj} with the characteristic function given in $(27)$, where the term associated with $n=2$ is absent as the topological Gauss-Bonnet invariant is used at second order in the curvature. This can be seen for instance also from the expression~\eqref{eq:hEGB} later.} In particular the characteristic function determining static spherically symmetric solutions in this case is a polynomial in $\psi$ at each finite order $n$. However, one may construct non-polynomial characteristic functions by a resummation of the non-polynomial curvature quasi-topological densities constructed in~\cite{Bueno:2025zaj} as $n \to \infty$, which resembles the mechanism of resummation of polynomial curvature quasi-topological densities in $d\geq 5$ necessary for obtaining regular black holes~\cite{Bueno:2024dgm}.
	
 We do not make any assumptions on the reduced theory here and in particular later when stating a Birkhoff theorem for generic $d$-dimensional gravities with second-order equations on $2+(d-2)$ warped-product backgrounds~\eqref{eq:Metric} with $q$ and $\varphi$ of the form~\eqref{eq:qabPhiSchwarzschildGauge1}--\eqref{eq:qabPhiSchwarzschildGauge2}. The discussions in~\cite{Colleaux:2017ibe,Colleaux:2019ckh,Bueno:2025zaj} are aimed  at emphasising the possibility of reducing $4$-dimensional non-polynomial higher-order densities to polynomial second-order densities in the spherically symmetric sector, which permits to view these theories as extensions of Lovelock or more generally polynomial CQTGs existing in $d\geq 5$, on  this particular class of $4$-dimensional backgrounds, whereas here we are interested in making a general statement about the possibility of generating the equations of motion of two-dimensional Horndeski theories on~\eqref{eq:qabPhiSchwarzschildGauge1}--\eqref{eq:qabPhiSchwarzschildGauge2} from the reduction of $d$-dimensional gravities, as well as in an interpretation of the corresponding onshell configurations as $d$-dimensional gravitational vacua.
\\

It should also be emphasised that the Weyl invariants appearing in the denominators of the covariant densities constructed above, similarly as for the constructions considered in~\cite{Colleaux:2017ibe,Colleaux:2019ckh,Bueno:2025zaj}, vanish on conformally flat backgrounds. In our subsequent  discussions we consider only onshell configurations $q$ and $\varphi$ of the  type~\eqref{eq:qabPhiSchwarzschildGauge1}--\eqref{eq:qabPhiSchwarzschildGauge2} corresponding to onshell configurations $g$ with non-vanishing Weyl tensor generically. This is motivated by our main goal of discussing two-dimensional Horndeski theories as reduced gravities in the context of black holes. In applications of the two-dimensional Horndeski framework to $d$-dimensional cosmology, one should first reduce the gravitational action on~\eqref{eq:Metric4d} and subsequently implement the cosmological ansatz, as~e.g.~in~\cite{Colleaux:2015yta,Colleaux:2019ckh,Bueno:2025zaj,Borissova:2026klg}.~\footnote{Similarly, the density $\mathcal{I}_{C^2} \mathcal{I}_{\hat{R}^2C} + 2\mathcal{I}_{\hat{R}^2} \mathcal{I}_{C^3}$ used in the denominators of the above examples of invariants reducing to $\mathcal{I}_\mathcal{R}$ and $\mathcal{I}_\psi$ vanishes when $q_{ab}$ and $\varphi$ are given by~\eqref{eq:qabPhiSchwarzschildGauge1f}--\eqref{eq:qabPhiSchwarzschildGauge2f}. See also~\cite{Bueno:2025zaj} for a construction with similar properties. However, when varying an action involving this particular realisation of invariants, the ansatz~\eqref{eq:qabPhiSchwarzschildGauge1f}--\eqref{eq:qabPhiSchwarzschildGauge2f} must in any case be inserted only after the reduction on a general background~\eqref{eq:Metric4d}.}

\section{Subset of 2D Horndeski theories from gravities $\mathcal{L}\qty(g^{\mu\nu},R_{\mu\nu\rho\sigma})$}\label{Sec:Local}

From the discussion in subsections~\ref{SecSubSub:LocalInvariants} and~\ref{SecSubSub:LocalNonPolynomialDensities} it follows that for the Horndeski actions~\eqref{eq:SHorndeskiRewritten} which can arise from the evaluation of a $d$-dimensional pure-cuvature Lagrangian density with second-order equations of motion on~\eqref{eq:Metric}, it must hold
\ba\label{eq:LLocalhiConstraint}
\mathcal{L}\big(g^{\mu\nu},R_{\mu\nu\rho\sigma}\big) \,\,\, \quad \rightarrow \,\,\, \quad {\mathbb{h}}_i(\varphi,\chi) \,\, = \,\, {\mathbb{h}}_i\qty(\psi)\,,
\ea
i.e., these Horndeski actions take the form
\ba\label{eq:SHorndeskiSubclass1}
S_{\text{Horndeski}}[q,\varphi] &=& \int \dd[2]{y} \sqrt{-q} \,\varphi^{d-2}\,\Bigg[{\mathbb{h}}_2(\psi) - {\mathbb{h}}_3(\psi) \frac{\Box \varphi}{\varphi} + \mathbb{h}_4(\psi)\mathcal{R} \nn\\
&+& 2  \mathbb{h}'_4(\psi)\qty[\qty(\frac{\Box \varphi}{\varphi})^2 - \frac{\nabla_a \nabla_b  \varphi}{\varphi} \frac{ \nabla^a \nabla^b \varphi}{\varphi}]\Bigg]\,.\,\,\,\,\,\quad 
\ea
The constraint~\eqref{eq:LLocalhiConstraint} must be understood prior to preforming partial integrations following the reduction. Partial integrations will generally spoil this particular functional dependence of the functions $\mathbb{h}_i$. In order to determine which physically inequivalent Horndeski theories can be generated from the reduction of a pure-curvature Lagrangian density, in the following we consider the structure of the functions $\alpha(\varphi,\chi)$ and $\beta(\varphi,\chi)$ entering the Horndeski equations of motion~\eqref{eq:Eab}. 

\subsection{Structure of the equations of motion}\label{SecSub:EOMLocal}

It will be convenient to perform the 
change of variables used already implicitly in~\eqref{eq:SLocal} and~\eqref{eq:SHorndeskiSubclass1},
\ba\label{eq:VariableChange}
(\varphi,\,\chi)\quad \,\, \to \,\,\quad  (\varphi,\,\psi) \,, \,\,\, \quad \quad 
(\partial_\varphi,\,\partial_\chi) \quad \,\, \to \,\, \quad \qty(\partial_\varphi -  \frac{2 \psi}{\varphi}\partial_\psi,\, -\frac{1}{\varphi^2}\partial_\psi)\,,
\ea
where the variable $\psi$ is defined in~\eqref{eq:Psi}. The functions $\alpha$ and $\beta$ defined in~\eqref{eq:AlphaBeta} for the subclass of Horndeski theories with action~\eqref{eq:SHorndeskiSubclass1} are then given by
\ba
\alpha(\varphi,\psi) &=&\varphi^{d-2} \mathbb{g}_1(\psi) + \varphi^{d-4} \,k\, \mathbb{g}_2(\psi)\,,\label{eq:AlphaLocal}\\
\beta(\varphi,\psi) &=& \varphi^{d-3}\mathbb{g}_3(\psi) + \varphi^{d-5} \,k\,\mathbb{g}_4(\psi)\,,\label{eq:BetaLocal}
\ea
where
\ba
\mathbb{g}_1(\psi) &=& \mathbb{h}_2 - (d-3)\psi \mathbb{h}_3 + 2 \psi^2 \mathbb{h}'_3 + 2(d-3)(d-2) \psi \mathbb{h}_4 - 4 (2d - 7)\psi^2 \mathbb{h}_4' + 8 \psi^3 \mathbb{h}_4''\,,\quad \label{eq:g1}\\
\mathbb{g}_2(\psi) &=&  (d-3)\mathbb{h}_3 - 2 \psi \mathbb{h}'_3 - 2(d-3)(d-2) \mathbb{h}_4 + 4 (2d - 7)\psi \mathbb{h}_4' - 8 \psi^2 \mathbb{h}_4''\,,\label{eq:g2}\\
\mathbb{g}_3(\psi) &=& \psi \mathbb{h}'_3  - (d-2)\mathbb{h}_4 - 2(d-5)\psi \mathbb{h}_4' + 4 \psi^2 \mathbb{h}_4''\,,\label{eq:g3} \quad \\
\mathbb{g}_4(\psi) &=& -\mathbb{h}_3' + 2(d-4)\mathbb{h}_4' - 4 \psi \mathbb{h}_4''\,. \quad \label{eq:g4}
\ea
These expressions are the first the main statement in this section. For any two-dimensional Horndeski theory arising from the reduction of a $d$-dimensional curvature Lagrangian density, the functions $\alpha(\varphi,\psi)$ and $\beta(\varphi,\psi)$ must be of the above form. This complements the discussion in subsection~\ref{SecSubSub:LocalNonPolynomialDensities} showing that for generic Horndeski theories relevant to the description of black holes with $\alpha(\varphi,\psi)$ and $\beta(\varphi,\psi)$ of the above form, it is possible to explicitly construct a $d$-dimensional curvature Lagrangian density whose reduction on~\eqref{eq:Metric} generates this theory. 

\subsection{Integrability condition}\label{SecSub:IntegrabilityLocal}

We will now discuss the  integrability condition~\eqref{eq:Integrability}
for the subset of Horndeski actions~\eqref{eq:SHorndeskiSubclass1} which can be generated from the reduction of a $d$-dimensional curvature Lagrangian density. From the discussion in subsection~\ref{SecSub:Birkhoff}, it follows that this condition is necessary and sufficient for the existence of solutions
\ba\label{eq:MetricToroidal}
g_{\mu\nu}(x)\dd{x}^\mu \dd{x}^\nu &=& -n(t,r)^2 f(t,r)\dd{t}^2 + \frac{\dd{r}^2}{f(t,r)} + r^2 \dd{\Sigma_{d-2}^2}
\ea
satisfying
\ba\label{eq:EOMIntegrability}
\beta \partial_t f \,\, = \,\, 0\,, \,\,\, \quad\quad  \beta \partial_r n \,\,= \,\, 0\,, \,\,\,\quad \quad \alpha + \beta \partial_r f \,\, = \,\, 0\,,
\ea
i.e., if the function $\beta$ is non-zero, then $f=f(r)$ and $n=n(t)$ whereby $n =1$ can be achieved by a redefinition of the coordinate $t$. The last equation in this case can be integrated to yield an algebraic equation for $f$ which amounts to the statement that the generating function $\Omega(\varphi,\chi)$ discussed in subsection~\ref{SecSub:SolvingIntegrableTheories} evaluates to an integration constant onshell, cf.~equations~\eqref{eq:OmegaEOM} and~\eqref{eq:OmegaOnshell}.\\

The integrability condition~\eqref{eq:Integrability} can be expressed in terms of the  variables~\eqref{eq:VariableChange} as
\ba\label{eq:IntegrabilityVariableChange}
-\frac{1}{\varphi^2} \partial_\psi \alpha - \partial_\varphi \beta + \frac{2\psi}{\varphi}\partial_\psi \beta &=& 0\,.
\ea
Therein inserting the expressions for $\alpha$ and $\beta$ in~\eqref{eq:AlphaLocal}--\eqref{eq:BetaLocal} leads to
\ba\label{eq:HiConstraint}
\mathbb{h}_2' - (d-3)\mathbb{h}_3 +  2 \psi \mathbb{h}_3' + (d-3)(d-2)\mathbb{h}_4 - 2 (2d -7) \psi \mathbb{h}_4' + 4 \psi^2 \mathbb{h}_4'' &=& 0\,.
\ea
Different from~\eqref{eq:IntegrabilityVariableChange}, this expression is not invariant under redefinitions of the functions $\mathbb{h}_i$ which amount to performing partial integrations in the reduced action. As emphasised previously, performing partial integrations in the Horndeski action will generally spoil the particular functional dependence $\mathbb{h}_i(\varphi,\chi)=\mathbb{h}_i(\psi)$. However, the above constraint is still meaningful when we refer to the functions $\mathbb{h}_i$ as those defined by the immediate evaluation of the $d$-dimensional Lagrangian density on~\eqref{eq:Metric} without performing partial integrations following this evaluation. This will allow us later to state the algebraic equation satisfied by the function $f$ in~\eqref{eq:MetricToroidal} in terms of functions which can be derived from that evaluated Lagrangian density.\\

The above constraint on the functions $\mathbb{h}_i$ can be written in terms of the functions $\mathbb{g}_i$ defined in~\eqref{eq:g1}--\eqref{eq:g4} as
\ba
\mathbb{g}_1' + (d-3)\mathbb{g}_3 - 2 \psi \mathbb{g}_3' &=& 0\,.
\ea
Its integrated version is
\ba\label{eq:giConstraint}
\mathbb{g}_1 + (d-3)\int \dd{\psi }\mathbb{g}_3 - 2 \int \dd{\psi} \psi \mathbb{g}_3' &=& \mathbb{g}_1 + (d-1)\int \dd{\psi }\mathbb{g}_3 - 2 \psi \mathbb{g}_3 \,\,=\,\,0
\ea
up to a constant. This constant will be irrelevant later, as it only amounts to a constant shift of the generating function $\Omega(\varphi,\chi)$ which can be absorbed by a  redefinition of the integration constant arising in the solution to the last equation of motion in~\eqref{eq:EOMIntegrability}.\\

We will now discuss separately the solutions of the form~\eqref{eq:MetricToroidal} to polynomial and non-polynomial CQTGs. Concretely, in subsection~\ref{SecSub:PolynomialLocalQTG} we first review statements about spherically symmetric solutions to polynomial CQTGs derived in~\cite{Bueno:2025qjk} whereby the action is assumed to be an analytic function of curvature invariants. Subsequently, in subsection~\ref{SecSub:NonPolynomialLocalQTG} we will discuss more generally solutions~\eqref{eq:MetricToroidal} to generic non-polynomial CQTGs.

\subsubsection{Polynomial curvature quasi-topological gravities}\label{SecSub:PolynomialLocalQTG}

In the following we will review statements on spherically symmetric solutions 
\ba\label{eq:MetricSpherical}
g_{\mu\nu}(x)\dd{x}^\mu \dd{x}^\nu &=& -n(t,r)^2 f(t,r)\dd{t}^2 + \frac{\dd{r}^2}{f(t,r)} + r^2 \dd{\Omega_{d-2}^2}
\ea
derived for polynomial CQTGs in~\cite{Bueno:2025qjk}. In this case $k=1$,  but the discussion can be straightforwardly generalised to $d-2$ dimensional compact spaces of non-zero constant sectional curvature $k$ other than the sphere, as its derivation depends only on the scalar $\psi$. We will take this into account when generalising these statements to non-polynomial CQTGs in the next subsection.
\\

If a $d$-dimensional curvature Lagrangian density with second-order equations of motion on~\eqref{eq:MetricSpherical} is a polynomial function of curvature invariants, the functions $\mathbb{h}_i$ in~\eqref{eq:SHorndeskiSubclass1} are constrained to satisfy~\cite{Bueno:2025qjk}
\ba\label{eq:hiPolynomial}
\mathbb{h}_2(\psi) \,\, = \,\, \qty(d-1)\mathbb{h} - 2 \psi\, \mathbb{h}'\,,\,\,\,\quad \quad \mathbb{h}_3(\psi) \,\, = \,\, 2 \mathbb{h}'\,,\,\,\,\quad \quad \mathbb{h}_4(\psi) \,\, = \,\, - \frac{1}{2}\psi^{(d-2)/2} \int \dd{\psi} \frac{\mathbb{h}'}{\psi^{d/2}} \,,\quad 
\ea
where the function $\mathbb{h}(\psi)$ can be deduced by evaluating the $d$-dimensional Lagrangian density on a single-function static spherically symmetric ansatz
\ba \label{eq:MetricSSS}
g_{\mu\nu}(x)\dd{x}^\mu \dd{x}^\nu &=& - f(r) \dd{t}^2 + \frac{\dd{r}^2}{f(r)} + r^2 \dd{\Omega_{d-2}^2}\,,
\ea
 in which case $\varphi  = r$, $\chi = f$ and 
\ba
\psi(r,f) &=& \frac{1-f}{r^2}\,.
\ea
The relevant formula is~\cite{Bueno:2025qjk}
\ba\label{eq:hPolynomial}
\mathbb{h}\qty(\psi(r,f)) &=& \frac{1}{d-1}\qty{\eval{\qty[\eval{\mathcal{L}}_{\eqref{eq:MetricSSS}}]}_{f' = 0,\, f'' = 0} -r\, \psi(r,f) \eval{\qty[\partial_{f'} \eval{\mathcal{L}}_{\eqref{eq:MetricSSS}}]}_{f' = 0}}\,.
\ea
In other words, the reduced actions~\eqref{eq:SHorndeskiSubclass1} for polynomial curvature gravities with second-order equations of motion on~\eqref{eq:MetricSpherical} can be parametrised in terms of a single one-variable function of $\psi$. 
Polynomial higher-curvature extensions of general relativity beyond standard Lanczos-Lovelock theories~\cite{Lanczos:1938sf,Lovelock:1970zsf,Lovelock:1971yv}
with second-order equations of motion on spherically symmetric  backgrounds~\cite{Oliva:2010eb,Myers:2010ru,Dehghani:2011vu,Cisterna:2017umf,Bueno:2019ltp,Bueno:2019ycr, Bueno:2022res,Moreno:2023rfl,Bueno:2025qjk},
to which we refer here as polynomial CQTGs, exist only in $d \geq 5$ dimensions~\cite{Bueno:2019ltp,Moreno:2023rfl}. From~\eqref{eq:hiPolynomial} it follows that 
\ba
\mathbb{h}_2' &=&\frac{1}{2}(d-3)\mathbb{h}_3 - \psi \mathbb{h}_3'\,,\label{eq:h2h3}\\
\mathbb{h}_3 &=& 
2(d-2)\mathbb{h}_4
- 4\psi \mathbb{h}_4'\,,\label{eq:h4h3}
\ea
which can be used to verify explicitly that these theories satisfy the integrability condition~\eqref{eq:HiConstraint}.~\footnote{See for instance also explicitly the constraint equation (32) in~\cite{Bueno:2024zsx} concerning the Horndeski functions $G_i(\varphi,\chi) \propto h_i(\varphi,\chi)$. This constraint implies the integrability condition~\eqref{eq:Integrabilityh}. Alternatively, one may directly verify~\eqref{eq:Integrabilityh} by using the expressions for the functions $G_i(\varphi,\chi)$ stated in~\cite{Bueno:2024zsx,Bueno:2024eig}.} Using the previous identities as well as~\eqref{eq:hiPolynomial} to compute the functions $\alpha$ and $\beta$ according to~\eqref{eq:AlphaLocal}--\eqref{eq:BetaLocal}, results in
\ba\label{eq:AlphaLocalPolynomial}
\alpha(\varphi,\psi) &=&
\varphi^{d-2} \big[(d-1)\mathbb{h}(\psi) - 2 \psi \mathbb{h}'(\psi)\big] \,,\\
\beta(\varphi,\psi)&=& 
- \varphi^{d-3}\mathbb{h}'(\psi)\,.\label{eq:BetaLocalPolynomial}
\ea
These expressions are manifestly of the form~\eqref{eq:AlphaLocal}--\eqref{eq:BetaLocal}.
The last equation of motion determining $f$ in~\eqref{eq:EOMIntegrability}  becomes
\ba
 (d-1)\mathbb{h}\qty(\frac{1-f}{r^2})- \qty[2 \qty(\frac{1-f}{r^2})+ \frac{f'}{r}]\mathbb{h}'\qty(\frac{1-f}{r^2})\,\,=\,\,0\,,
\ea
where $\psi(r,f)= (1-f)/r^2$ has been used.
This equation can be integrated to obtain the algebraic equation characterising spherically symmetric solutions to polynomial CQTGs~\cite{Bueno:2025qjk}\,,
\ba\label{eq:EqMasterLocalPolynomialQTG}
\mathbb{h}\qty(\frac{1-f}{r^2}) &=& \frac{2M}{r^{d-1}}\,,
\ea
where $M$ is an integration constant related to the ADM mass of the solution. \\

In summary, and for later comparison with non-polynomial CQTGs, we can state the following result derived in~\cite{Bueno:2025qjk} in adapted formulation and notation.\\

{{\bf Theorem~\cite{Bueno:2025qjk}:} {\it The unique spherically symmetric solutions of a $d\geq 5$ dimensional gravitational theory $\mathcal{L}\qty(g^{\mu\nu},R_{\mu\nu\rho\sigma})$ constructed from analytic functions of curvature invariants without covariant derivatives and with second-order equations of motion on~\eqref{eq:MetricSpherical}, are given by
\ba
\mathbb{h}'\qty(\psi)\partial_t f \,\, = \,\, 0\,, \,\,\,\quad \quad \mathbb{h}'\qty(\psi)\partial_r n \,\, = \,\, 0\,, \,\,\, \quad \quad \derivative{}{r}\qty[r^{d-1}\mathbb{h}(\psi)] \,\,=\,\,0 \,,
\ea
where $\mathbb{h}$ is a theory-dependent analytic function given by~\eqref{eq:hPolynomial} and $\psi = (1-f)/r^2$.}}\\

{\it Example:}  As a prototype example of a polynomial CQTG, we can consider the Einstein-Gauss-Bonnet (EGB) action~\cite{Lovelock:1970zsf,Lovelock:1971yv} 
\ba\label{eq:SEGB}
S_{\text{EGB}}[g] &=& \int \dd[d]{x}\sqrt{-g}\,\qty[R + \gamma \,\mathcal{G}]\,, \,\,\, \quad \quad \mathcal{G} \,\, = \,\,  R^2 - 4 R_{\mu\nu} R^{\mu\nu} + R_{\mu\nu\rho\sigma}R^{\mu\nu\rho\sigma}\,,
\ea
where $R$ is the Ricci scalar and $\mathcal{G}$ the Gauss-Bonnet invariant. Their evaluation on~\eqref{eq:Metric} is given by
\ba
\eval{R}_{\eqref{eq:Metric}} &=&  (d-3)(d-2)\psi - 2\qty(d-2)\frac{\Box \varphi}{\varphi} +\mathcal{R}  \,,\\
\eval{\mathcal{G}}_{\eqref{eq:Metric}} &=& \qty(d-3)\qty(d-2)\Bigg[ (d-4)(d-5) \psi^2 - 4 (d-4)\psi \frac{\Box \varphi}{\varphi}+ 2  \psi \mathcal{R} \nn\\
&+&  4 \qty[\qty(\frac{\Box \varphi}{\varphi})^2 - \frac{\nabla_a \nabla_b \varphi}{\varphi} \frac{\nabla^a \nabla^b \varphi}{\varphi} ] \Bigg ]\,.
\ea
Comparing with~\eqref{eq:SHorndeskiSubclass1}, we can identify the functions $\mathbb{h}_i$ as
\ba
\mathbb{h}_{2}(\psi) &=&  (d-3)(d-2) \qty[\psi + \gamma(d-4)(d-5)\psi^2 ]\,,\\
\mathbb{h}_{3}(\psi)&=&   (d-2)\qty[2 + 4 \gamma (d-4)(d-3) \psi ]\,,\\
\mathbb{h}_{4}(\psi) &=& 1 + 2 \gamma(d-3)(d-2)\psi\,.
\ea
One may verify that these satisfy the relations~\eqref{eq:hiPolynomial} with
\ba\label{eq:hEGB}
\mathbb{h}_{\text{EGB}}(\psi) &=& \frac{1}{d-1}\qty[\mathbb{h}_2 + \psi \mathbb{h}_3] \,\,=\,\, (d-2)\qty[\psi + \gamma(d-4)(d-3) \psi^2]\,,
\ea
which in particular confirms  the integrability condition~\eqref{eq:HiConstraint}.
The functions $\alpha$ and $\beta$ computed according to~\eqref{eq:AlphaLocalPolynomial}--\eqref{eq:BetaLocalPolynomial} are
\ba
\alpha_{\text{EGB}}(\varphi,\psi) &=& \qty(d-3)(d-2) \qty[\psi + \gamma (d-4)(d-5)\psi^2] \varphi^{d-2}\,,\label{eq:AlphaEGB}\\
\beta_{\text{EGB}}(\varphi,\psi) &=& -(d-2)\qty[1+ 2\gamma  (d-4)(d-3)\psi] \varphi^{d-3}\,.\label{eq:BetaEGB}
\ea
The terms proportional to the EGB coupling $\gamma$ in~\eqref{eq:hEGB} and~\eqref{eq:AlphaEGB}--\eqref{eq:BetaEGB} involve a factor $(d-4)$ indicating that there is no contribution to the equations of motion coming from the Gauss-Bonnet invariant in the action~\eqref{eq:SEGB} for $d=4$. It is well-known that this term is a topological invariant in $d=4$.~\footnote{See, however, \cite{Glavan:2019inb,Fernandes:2020rpa,Konoplya:2020qqh,Ghosh:2020vpc,Ghosh:2020syx,Konoplya:2020juj,Kumar:2020owy,Kumar:2020uyz,Kumar:2020bqf,Konoplya:2020cbv,Malafarina:2020pvl,Yang:2020jno,Feng:2020duo, Gurses:2020ofy,Gurses:2020rxb,Arrechea:2020evj,Arrechea:2020gjw,Bonifacio:2020vbk,Ai:2020peo,Mahapatra:2020rds,Hohmann:2020cor,Cao:2021nng} for discussions on defining EGB gravity in $d=4$ and~\cite{Fernandes:2022zrq} for a review.} From~\eqref{eq:hEGB}, by using the algebraic equation~\eqref{eq:EqMasterLocalPolynomialQTG}, we obtain the familiar static spherically symmetric Boulware-Deser black-hole metric~\cite{Boulware:1985wk}
\ba
f_{\text{EGB}}(r) &=& 1 + \frac{r^2}{2\gamma (d-4)(d-3) }\Bigg[1 \pm \sqrt{1 + \frac{8  \gamma M(d-4)(d-3)}{ r^{d-1}} }\,\Bigg]\,.
\ea

\subsubsection{Non-polynomial curvature quasi-topological gravities}\label{SecSub:NonPolynomialLocalQTG}

For a generic $d$-dimensional non-polynomial curvature Lagrangian density with second-order equations of motion on~\eqref{eq:MetricToroidal}, the functions $\mathbb{h}_i(\psi)$ in the reduced action~\eqref{eq:SHorndeskiSubclass1} can be generic. In this case one may distinguish between theories satisfying the integrability condition~\eqref{eq:IntegrabilityVariableChange} and hence the constraint~\eqref{eq:HiConstraint} on the functions $\mathbb{h}_i$, and those that do not satisfy this condition, as observed in~\cite{Borissova:2026wmn}. We refer here to the former theories as non-polynomial CQTGs. From the discussion in subsection~\ref{SecSub:Birkhoff} it follows that non-polynomial curvature gravities with second-order field equations on~\eqref{eq:MetricToroidal} which are not of quasi-topological type admit solutions of the form~\eqref{eq:MetricToroidal} with $\partial_r n \neq 0$. We do not discuss such theories here.\\

To solve the integrability condition~\eqref{eq:Integrability} expressed in terms of the variables $\varphi$ and $\psi$ in~\eqref{eq:IntegrabilityVariableChange} for generic curvature theories with second-order equations of motion on~\eqref{eq:MetricToroidal}, we will resort to the construction of a characteristic function $\Omega(\varphi,\psi)$ as discussed in subsection~\ref{SecSub:SolvingIntegrableTheories}. The defining relations~\eqref{eq:IntegrabilityOmega} expressed in terms of $\varphi$ and $\psi$ are
\ba\label{eq:IntegrabilityOmegaVariableChange}
\alpha(\varphi,\psi) \,\,=\,\, \partial_\varphi \Omega(\varphi,\psi) - \frac{2\psi}{\varphi} \partial_\psi \Omega(\varphi,\psi)\,, \,\,\,\quad \quad \beta(\varphi,\psi) \,\, =\,\, - \frac{1}{\varphi^2}\partial_\psi \Omega(\varphi,\psi)\,.
\ea
The last equation can be integrated to obtain
\ba
\Omega(\varphi,\psi) &=& -\varphi^2 \int \dd{\psi} \beta(\varphi,\psi) + F(\varphi)\,,
\ea
where $F$ is an apriori arbitrary function of $\varphi$. The first equation in~\eqref{eq:IntegrabilityOmegaVariableChange} imposes however $F' = 0$, and thus without loss of generality we may set $F =0$. Any other constant choice of $F$ can be absorbed into the integration constant arising when solving the remaining equation of motion in~\eqref{eq:EOMIntegrability} in the form $\Omega(r,f)=2M$ onshell. Therefrom, by using the expression  for the function $\beta $ in~\eqref{eq:BetaLocal}, we compute
\ba\label{eq:OmegaLocal}
\Omega(\varphi,\psi) &=& - \varphi^2 \int \dd{\psi}\beta(\varphi,\psi) \,\,= \,\,-  \varphi^{d-1}\int \dd{\psi} \mathbb{g}_3 - \varphi^{d-3}\, k \int \dd{\psi}\mathbb{g}_4\nn\\
&=&  \frac{1}{d-1} \qty[\mathbb{g}_1 - 2 \psi \mathbb{g}_3] \varphi^{d-1} +k\, \qty[\mathbb{h}_3 - 2(d-2)\mathbb{h}_4 + 4 \psi \mathbb{h}_4']\varphi^{d-3}\nn\\
& \equiv &
 \qty[\mathbb{h}- \frac{d-2}{d-1}\psi\mathbb{g}] \varphi^{d-1} + k\,\mathbb{g} \varphi^{d-3}\,,
\ea
where in the second line~\eqref{eq:g4} and~\eqref{eq:giConstraint} were used, and moreover we have defined
\ba
\mathbb{h}(\psi) &=& \frac{1}{d-1}\qty[\mathbb{h}_2 + \psi \mathbb{h}_3] \,,\label{eq:hDef}\\
\mathbb{g}(\psi) &=& \mathbb{h}_3 - 2(d-2)\mathbb{h}_4 + 4 \psi \mathbb{h}_4'\label{eq:gDef}\,.
\ea
The function $\mathbb{h}$ coincides with the previously defined function for  polynomial CQTGs~\eqref{eq:hiPolynomial}, and for these theories it holds $\mathbb{g} =0$ according to~\eqref{eq:h4h3}.\\

With the above definitions, the functions $\alpha $ and $\beta$ in~\eqref{eq:IntegrabilityOmegaVariableChange} for non-polynomial CQTGs can be written as
\ba
\alpha(\varphi,\psi) &=& -\frac{1}{d-1}\varphi^{d-2}\qty[(d-3)(d-2)\psi \mathbb{g}- 2 (d-2)\psi^2 \mathbb{g}'  - (d-1)^2 \mathbb{h} + 2(d-1)\psi \mathbb{h'}] \nn\\
&+& \varphi^{d-4} \,k \,\qty[(d-3)\mathbb{g} - 2 \psi \mathbb{g}']\,,\quad \quad\\
\beta(\varphi,\psi) &=& \frac{1}{d-1}\varphi^{d-3} \qty[(d-2)\mathbb{g} + (d-2)\psi \mathbb{g}' - (d-1)\mathbb{h}'] - \varphi^{d-5} \,k\,\mathbb{g}'\label{eq:Betagh}\,.
\ea
One may verify that these reduce to~\eqref{eq:AlphaLocalPolynomial}--\eqref{eq:BetaLocalPolynomial} when $\mathbb{g}=0$.\\

Finally, as described in subsection~\ref{SecSub:SolvingIntegrableTheories}, the remaining equation of motion in~\eqref{eq:EOMIntegrability} can be integrated to yield the following algebraic equation determining $f$,
\ba\label{eq:EqMasterLocalNonPolynomialQTG}
 r^{d-1}\qty[\mathbb{h}\qty(\frac{k-f}{r^2})- \frac{d-2}{d-1}\qty(\frac{k-f}{r^2})\mathbb{g}\qty(\frac{k-f}{r^2})]  + r^{d-3}\,k\,\mathbb{g}\qty(\frac{k-f}{r^2})  \,\, = \,\, 2M\,,
\ea
where we have used $\psi = (k-f)/r^2$. This is our key result generalising statements on polynomial CQTGs in $d \geq 5$~\cite{Bueno:2025qjk}  and statements on non-polynomial CQTGs in $d=4$~\cite{Borissova:2026wmn} to non-polynomial CQTGs in $d\geq 4$.\\

We can collect these results in the following theorem.\\

{\bf{Theorem:}} {\it Let $\mathcal{L}\qty(g^{\mu\nu},R_{\mu\nu\rho\sigma})$ be a $d$-dimensional gravitational theory constructed from curvature invariants without covariant derivatives and possessing second-order equations of motion  on~\eqref{eq:MetricToroidal}. Assume the two-dimensional Horndeski theory obtained from its reduction on these backgrounds satisfies the integrability condition~\eqref{eq:Integrability}, i.e., $\mathcal{L}$ is of quasi-topological type. Then the unique solutions satisfy
\ba\label{eq:EOMTheorem}
\beta \partial_t f \,\,=\,\, 0\,,\,\,\,\quad \quad  \beta \partial_r n \,\,=\,\, 0\,,\,\,\,\quad \quad 
 \derivative{}{r}\qty[ r^{d-1}\qty[\mathbb{h}\qty(\psi)- \frac{d-2}{d-1}\psi\mathbb{g}\qty(\psi)]  + r^{d-3}\,k\,\mathbb{g}\qty(\psi) ] \,\,=\,\,0\,,\quad \,\,
\ea
where $\beta$ is given in terms of two theory-dependent functions $\mathbb{h}$ and $\mathbb{g}$ as in~\eqref{eq:Betagh} and $\psi = (k-f)/r^2$. Moreover the functions $\mathbb{h}$ and $\mathbb{g}$ can be computed from the reduced Lagrangian density evaluated on a single-function static ansatz
\ba\label{eq:MetricStatick}
g_{\mu\nu}(x)\dd{x}^\mu \dd{x}^\nu &=& - f(r)\dd{t}^2 + \frac{\dd{r}^2}{f(r)} + r^2 \dd{\Sigma_{d-2}^2}\,,
\ea
concretely} $\mathcal{L}_f \equiv\qty[ \varphi^{2-d} \mathcal{L}_{\text{\text{Horndeski}}}]_{\eqref{eq:MetricStatick}}$ {\it defined through~\eqref{eq:SHorndeskiRewritten},
as follows,
\ba
\mathbb{h}\qty(\psi(r,f)) &=& \frac{1}{d-1}\qty[\eval{\mathcal{L}_f}_{f' = 0,\, f'' = 0} -r\, \psi(r,f) \eval{\partial_{f'} \mathcal{L}_f }_{f' = 0}]\,,\quad \label{eq:hh}\\
\mathbb{g}\qty(\psi(r,f)) &=& - r\eval{\partial_{f'} \mathcal{L}_f }_{f' = 0} + 2\,r^2 \,\psi(r,f) \,\partial_{f'}^2\mathcal{L}_f  \label{eq:gg}   + 2(d-2)\,\partial_{f''}\mathcal{L}_f  \,,\quad 
\ea
}
The last formula in~\eqref{eq:EOMTheorem} is understood with the replacement $f \to f(r)$ in $\psi$ before taking the total derivative with respect to $r$.\\

{\it Proof:} Symmetric criticality implies that any $d$-dimensional gravitational action with second-order equations of motion on~\eqref{eq:MetricToroidal} reduces to a Horndeski theory for the two-dimensional metric $q_{ab}(y)$ and scalar field $\varphi(y)$ given in~\eqref{eq:qabPhiSchwarzschildGauge1}--\eqref{eq:qabPhiSchwarzschildGauge2}. If this Horndeski theory satisfies the integrability condition~\eqref{eq:Integrability}, then the equations of motion~\eqref{eq:Eab} for the ansatz~\eqref{eq:qabPhiSchwarzschildGauge1}--\eqref{eq:qabPhiSchwarzschildGauge2} are explicitly~\eqref{eq:Eqft}--\eqref{eq:EqLast}. Equations~\eqref{eq:Eqft} and~\eqref{eq:Eqnr} are identical to the first two equations in~\eqref{eq:EOMTheorem}, whereas equation~\eqref{eq:EqLast} is equivalent to the last equation in~\eqref{eq:EOMTheorem} written in terms of the generating function $\Omega(\varphi,\chi)$ defined in~\eqref{eq:IntegrabilityOmega} --- see~\eqref{eq:IntegrabilityOmegaVariableChange} in the variables $\varphi$ and $\psi$ --- according to~\eqref{eq:OmegaEOM}, with $\varphi =r$ and $\chi = f$ onshell, i.e., $\psi = (k-f)/r^2$ onshell. We have seen that this function is uniquely determined up to an additive constant which is irrelevant when taking the total derivative with respect to $r$ in~\eqref{eq:EOMTheorem}. \\

That $\Omega(r,\psi(r,f))$ for two-dimensional Horndeski theories arising from the reduction of a $d$-dimensional gravitational action constructed from curvature invariants without covariant derivatives with second-order equations of motion on~\eqref{eq:MetricToroidal} can be expressed as in the square brackets in~\eqref{eq:EOMTheorem}, was the content of the discussion in subsection~\ref{SecSubSub:LocalInvariants} and the derivativion presented in this section leading to~\eqref{eq:OmegaLocal} with functions $\mathbb{h}$ and $\mathbb{g}$ defined in~\eqref{eq:hDef} and~\eqref{eq:gDef}. 
It only remains to show that these functions can be expressed as in~\eqref{eq:hh}--\eqref{eq:gg}. To that end, we note that the Riemann scalars in~\eqref{eq:BuildingBlocks} evaluated for the metric~\eqref{eq:MetricStatick} are given by
\ba
\eval{\qty{\mathcal{R}\,,\, \frac{\Box \varphi}{\varphi}\,,\, \frac{ \nabla_a \nabla_b \varphi}{\varphi} \frac{ \nabla^a \nabla^b \varphi}{\varphi} \,,\,\psi}}_{\eqref{eq:MetricStatick}} & = & \qty{- f''\,,\, \frac{f'}{r} \,,\, 	\frac{f'^2}{2 r^2 } \,,\, \frac{k-f}{r^2}}\,.
\ea
Therefore the reduced Lagrangian density evaluated on~\eqref{eq:MetricStatick} is given by~\footnote{Here we use that in fact $\mathcal{L}_{\text{Horndeski}}$ as given in~\eqref{eq:SHorndeskiSubclass1} is the most general functional which can arise from the evaluation of a curvature gravitational action with second-order equations of motion on~\eqref{eq:Metric}.}
\ba
\mathcal{L}_f &=& \mathbb{h}_2(\psi(r,f)) - \mathbb{h}_3(\psi(r,f))\frac{f'}{r} - \mathbb{h}_4(\psi(r,f))f'' + \mathbb{h}_4'(\psi(r,f)) \frac{f'^2}{ r^2} \,.
\ea
Therefrom $\mathbb{h}$ and $\mathbb{g}$ defined in~\eqref{eq:hDef} and~\eqref{eq:gDef} can be straightforwardly expressed in the form~\eqref{eq:hh}--\eqref{eq:gg}.
\qed
\\

{\it Example:} Let us consider exemplarily what is the subset of integrable dilaton theories of the specific form~\eqref{eq:SDilaton} and correspondingly solutions~\eqref{eq:MetricToroidal} which can be generated from the reduction of a $d$-dimensional pure-curvature Lagrangian density with second-order equations of motion on~\eqref{eq:MetricStatic}. In terms of the variables~\eqref{eq:VariableChange}, the generating functions $\alpha_\text{D}$ and $\beta_{\text{D}}$ in~\eqref{eq:AlphaDilaton}--\eqref{eq:BetaDilaton} are given by
\ba
\alpha_{\text{D}}(\varphi,\psi) &=&  - \mathcal{V} - 2 \qty(k- \varphi^2 \psi) D''\,,\label{eq:AlphaDilatonVariableChange}\\
\beta_{\text{D}}(\varphi,\psi) &=& - D' \label{eq:BetaDilatonVariableChange}\,.
\ea
The integrability condition~\eqref{eq:Integrability} requires $D'' = - \partial_\chi \mathcal{V} =  \partial_\psi \mathcal{V} / \varphi^2$. In particular $\mathcal{V}$ must be of the form
\ba
\mathcal{V}(\varphi,\psi) &=& \varphi^2 D'' \psi + \nu'(\varphi)\,,
\ea
where $\nu' $ is an a priori arbitrary function of $\varphi$.  Thus the functions~\eqref{eq:AlphaDilatonVariableChange}--\eqref{eq:BetaDilatonVariableChange} for integrable dilaton theories~\eqref{eq:SDilaton} can be expressed as 
\ba
\alpha_{\text{D}}(\varphi,\psi) &=& \varphi^2 D'' \psi  - \qty[\nu'  + 2 k D''] \,,\\
\beta_{\text{D}}(\varphi,\psi) &=& -D'\,.
\ea
The algebraic equation for the metric function $f$ in~\eqref{eq:MetricStatick} can be deduced by applying the derivation in terms of a characteristic function $\Omega_{\text{D}}(\varphi,\psi)$ as described in subsection~\ref{SecSub:SolvingIntegrableTheories}, cf.~equation~\eqref{eq:IntegrabilityOmega} leading to~\eqref{eq:OmegaOnshell}, and in the previous subsection, cf.~concretely equation~\eqref{eq:IntegrabilityOmegaVariableChange} in terms of the variables $\varphi$ and $\psi$. Hence we compute
\ba
\Omega_{\text{D}}(\varphi,\psi) &=& - \varphi^2 \int \dd{\psi}\beta_{\text{D}}(\varphi,\psi) \,\,=\,\, \varphi^2 D' \psi+ F(\varphi) \,.
\ea
Satisfying the defining equation for $\alpha_{\text{D}}$ in~\eqref{eq:IntegrabilityOmegaVariableChange} requires
\ba
\alpha_{\text{D}}(\varphi,\psi) &=& \partial_\varphi \Omega_{\text{D}} - \frac{2\psi}{\varphi}\partial_\psi \Omega_{\text{D}} 
\,\,=\,\, \varphi^2 D'' \psi + \partial_\varphi F \,\,\,\quad \,\,\, \Rightarrow \,\,\, \quad \,\,\, F \,\,=\,\, - \qty[\nu + 2 k D']\,,\quad 
\ea
up to an irrelevant integration constant. The algebraic equation $\Omega(r,\psi(r,f)) = 2M$ then becomes
\ba\label{eq:EqMasterDilatonIntegrable}
r^2 D' \psi(r,f)  - \qty[\nu + 2 kD'] \,\,=\,\, 2M \,\,\,\quad \,\,\,\Leftrightarrow \,\,\,\quad \,\,\, f_{\text{D}}(r) \,\,=\,\, -k - \frac{2M}{D'} - \frac{ \nu}{D'}\,,
\ea
where we assume that $D'$ and thus the function $\beta_{\text{D}}$ is non-zero.
However, for the above algebraic equation and accordingly solution for $f$ to arise from the reduction of a $d$-dimensional pure-curvature Lagrangian density, we must require the functional form of $\Omega(\varphi,\psi)$ specified in~\eqref{eq:OmegaLocal} which lead to the algebraic equation~\eqref{eq:EqMasterLocalNonPolynomialQTG}. In other words, only those integrable dilaton theories for which~\eqref{eq:EqMasterDilatonIntegrable} can be written in the form~\eqref{eq:EqMasterLocalNonPolynomialQTG} for some functions $\mathbb{h}$ and $\mathbb{g}$ of $\psi$ can arise from the reduction of a $d$-dimensional Lagrangian density built only from curvature invariants without covariant derivatives. We see that naively there are two possibilities. The first one is
\ba
r^2 D' \,\, \propto \,\, r^{d-1} \,\,\, \quad \Rightarrow \,\,\,\quad 
\begin{cases}
 -\qty[\nu + 2k D' ]\,\,\propto\,\, r^{d-3}\,,\,\,\, \mathbb{g}\, \,=\,\,\text{const}\,\,\neq \,\,0\,, \,\,\, \mathbb{h} \,\,\propto \,\,\psi \,\,\, \text{or} \,\,\,\mathbb{h} \,\, = \,\, 0\,,\quad \\
\nu \,\,= \,\,-2kD'\,,\,\,\, \mathbb{g} \,\,=\,\,0 \,,\,\,\,\mathbb{h} \,\,\propto \,\,\psi \,. \quad \\
\end{cases}
\ea
The second one is
\ba
r^2 D' \,\, \propto \,\, r^{d-3} \,\,\, \quad \Rightarrow \,\,\,\quad 
\begin{cases}
	\mathbb{g} \,\,\propto \,\,\psi \,, \,\,\, -\qty[\nu + 2 k D'] \,\,\propto \,\, r^{d-1}\,,\,\,\, \mathbb{h} \,\,=\,\, \text{const} + \frac{d-2}{d-1}\psi \mathbb{g}\,,\\
		\mathbb{g} \,\,\propto \,\,\psi \,, \,\,\, \nu \,\,=\,\, 2 k D'\,,\,\,\, \mathbb{h} \,\,=\,\,  \frac{d-2}{d-1}\psi \mathbb{g}\,.\\
\end{cases}
\ea
In any of these subcases by a rescaling of the integration constant $M$ one may cast the solution for $f$ into the form
\ba
f_{\text{D}}(r) \,\, =\,\, k + c - \frac{2M}{r^{d-3}} \,,
\ea
where $c$ is a constant depending on the particular on the subcase.
This just a topological generalisation of the $d$-dimensional Schwarzschild-Tangherlini metric~\cite{Tangherlini:1963bw}. Therefore, while there exist infinitely many inequivalent generally covariant $d$-dimensional actions built from curvature invariants whose reduction yields an integrable dilaton theory of the form~\eqref{eq:SDilaton}, these dilaton theories for the considered type of ansatz only produce the above solution. This restricted solution space is on the one hand a consequence of the assumption that the non-minimal coupling $D$ in the dilaton action~\eqref{eq:SDilaton} depends only on $\varphi$. On the other hand, we have here explicitly imposed the requirement that the integrable dilaton theory arises from a $d$-dimensional gravitational theory built only from curvature invariants without covariant derivatives --- once curvature-derivative invariants are allowed in the gravitational action, one may generate the equations of motion on~\eqref{eq:qabPhiSchwarzschildGauge1f}--\eqref{eq:qabPhiSchwarzschildGauge2f} of any desired integrable two-dimensional dilaton theory~\eqref{eq:SDilaton}.~\footnote{In fact, by constructing $d$-dimensional generally covariant densities $\mathcal{I}_\mathcal{R}$ and $\mathcal{I}_\chi$ which upon reduction on~\eqref{eq:Metric} reduce to $\mathcal{R}$ and $\chi$,  one may generate more general two-dimensional dilaton theories of higher order in the curvature and in derivatives of the scalar field.}

\section{All 2D Horndeski theories from gravities $\mathcal{L}\qty(g^{\mu\nu},R_{\mu\nu\rho\sigma},\nabla_\mu)$}\label{Sec:QuasiLocal}

The discussion in subsection~\ref{SecSubSub:QuasiLocalInvariants} and~\ref{SecSubSub:QuasiLocalNonPolynomialDensities} shows that for the Horndeski actions~\eqref{eq:SHorndeskiRewritten} which can arise from the reduction of a generic $d$-dimensional Lagrangian density involving curvature invariants as well as curvature-derivative invariants, and possessing second-order equations of motion on~\eqref{eq:Metric}, the functions $\mathbb{h}_i(\varphi,\chi)$ defined in~\eqref{eq:hhiDef} can be generic, i.e.,
\ba\label{eq:LQuasiLocalNohiConstraint}
\mathcal{L}\big(g^{\mu\nu},R_{\mu\nu\rho\sigma}, \nabla_\mu \big) \,\,\, \quad \rightarrow \,\,\, \quad {\mathbb{h}}_i(\varphi,\chi) \,\, = \,\, \text{generic}\,,
\ea
and in particular generic two-dimensional Horndeski theories relevant to the description of black holes can be generated in this way. 

\subsection{Structure of the equations of motion}\label{SecSub:EOMQuasiLocal}

The discussion about the structure of the equations of motion on~\eqref{eq:Metric} for such theories thus amounts to a discussion of the general  Horndeski equations of motion~\eqref{eq:Eab} with $\alpha(\varphi,\chi)$ and $\beta(\varphi,\chi)$ defined in~\eqref{eq:AlphaBeta} and rewritten optionally  in terms of the functions $\mathbb{h}_i(\varphi,\chi)$ in~\eqref{eq:hhiDef} as follows,
\ba
\alpha(\varphi,\chi) &=& \varphi^{d-2} \mathbb{g}_1(\varphi,\chi) + \varphi^{d-3} \mathbb{g}_2(\varphi,\chi)  + \varphi^{d-4} \mathbb{g}_3(\varphi,\chi) \,,\\
\beta(\varphi,\chi) &=& \varphi^{d-2} \mathbb{g}_4(\varphi,\chi) + \varphi^{d-3} \mathbb{g}_5(\varphi,\chi)  \,,
\ea
where
\ba
\mathbb{g}_1(\varphi,\chi) &=& \mathbb{h}_2 - 2 \chi \partial_\varphi^2 \mathbb{h}_4\,,\\
\mathbb{g}_2(\varphi,\chi) &=& \chi \partial_\varphi \qty(\mathbb{h}_3 - 4 (d-2)\mathbb{h}_4)\,,\\
\mathbb{g}_3(\varphi,\chi) &=& (d-3)\chi \qty(\mathbb{h}_3 - 2 (d-2)\mathbb{h}_4) \,,\\
\mathbb{g}_4(\varphi,\chi) &=& - \partial_\varphi \qty(\mathbb{h}_4 + 2\chi \partial_\chi \mathbb{h}_4)\,,\\
\mathbb{g}_5(\varphi,\chi) &=& -(d-2)\mathbb{h}_4 + \chi \partial_\chi \qty(\mathbb{h}_3 - 2(d-2)\mathbb{h}_4)\,.
\ea
The variable change~\eqref{eq:VariableChange} at this point is not particularly convenient.

\subsection{Integrability condition}\label{SecSub:IntegrabilityQuasiLocal}

The integrability condition~\eqref{eq:Integrability} can be written in terms of the functions $\mathbb{h}_i$ in~\eqref{eq:hhiDef} which determine the reduced action~\eqref{eq:SHorndeskiRewritten} as
\ba
(d-3)\mathbb{h}_3 - (d-3)(d-2)\mathbb{h}_4 + \varphi^2 \partial_\chi \mathbb{h}_2 + \varphi \partial_\varphi \mathbb{h}_3 - 2 (d-2) \varphi \partial_\varphi \mathbb{h}_4 - \varphi^2 \partial_\varphi^2\mathbb{h}_4 &=& 0\,.\label{eq:hhiIntegrabilityQuasiLocal}
\ea
As before, this representation of the integrability condition in terms of functions $\mathbb{h}_i$ is not invariant under partial integrations performed in the two-dimensional Horndeski action. However, this equation is still meaningful if we define the particular choice of representation of the functions $\mathbb{h}_i$ as the one arising from the  immediate evaluation of the $d$-dimensional Lagrangian density $\mathcal{L}$ on the ansatz~\eqref{eq:Metric} without performing partial integrations following this evaluation. This representation is the one that allows us  to state the algebraic equation satisfied by $f$ in terms of functions which can be derived from the evaluated Lagrangian density.~\footnote{As a cross check, one may verify that changing variables according to~\eqref{eq:VariableChange} in~\eqref{eq:hhiIntegrabilityQuasiLocal} and requiring that $\partial_\varphi \mathbb{h}_i(\varphi,\psi) = 0$ reproduces the previously stated special subcase of the integrability condition for CQTGs in~\eqref{eq:HiConstraint}.}\\

The above constraint can be expressed in terms of the function $\mathbb{f}$ defined as
\ba
\mathbb{f} &=&  \qty(\mathbb{h}_3 - (d-2)\mathbb{h}_4 - \varphi \partial_\varphi \mathbb{h}_4) \,\,\, \quad \,\,\, \Rightarrow \,\,\, \quad \,\,\,  \mathbb{f} \,\,=\,\, -\frac{1}{d-3}\qty[\varphi^2 \partial_\chi \mathbb{h}_2 + \varphi \partial_\varphi\mathbb{f}] \,. 
\ea
One may integrate this constraint twice to write
\ba
\int \dd{\chi} \mathbb{f} &=&   - \varphi^{3-d} \qty[\int \dd{\varphi} \varphi^{d-2}  \mathbb{h}_2
 ] \,,
\label{eq:IntChif}
\ea
where we have set any integration functions 
to zero as we are interested only in stating an expression for the characteristic function $\Omega(\varphi,\chi)$ which determines $f$  in terms of the functions $\mathbb{h}_i$ defined as those arising from the evaluation of the $d$-dimensional Lagrangian.

\subsubsection{All quasi-topological gravities}\label{SecSub:AllQTG}

We now repeat the discussion about the derivation of the algebraic equation for $f $ in terms of the characteristic function $\Omega(\varphi,\chi)$ satisfying the defining relations~\eqref{eq:IntegrabilityOmega}. As described in subsection~\ref{SecSub:SolvingIntegrableTheories}, we compute
\ba\label{eq:OmegaQuasiLocal}
\Omega(\varphi,\chi) &=& \int \dd{\chi} \beta(\varphi,\chi) \nn\\
&=&  \chi \varphi^{d-3}\qty[\mathbb{h}_3 - 2(d-2) \mathbb{h}_4 -2 \varphi\partial_\varphi \mathbb{h}_4] 
- \varphi^{d-3} \int \dd{\chi} \qty[\mathbb{h}_3 - (d-2) \mathbb{h}_4 - \varphi\partial_\varphi \mathbb{h}_4]\nn\\
&\equiv & \chi  \varphi^{d-3} \mathbb{G} - \varphi^{d-3} \int \dd{\chi} \mathbb{f} \,\, = \,\, \chi  \varphi^{d-3} \mathbb{G}  +\int \dd{\varphi} \varphi^{d-2} \, \mathbb{H}\,,
\ea
 where in the last step we have used~\eqref{eq:IntChif} and moreover have defined
\ba
\mathbb{G}(\varphi,\chi) &=&  \mathbb{h}_3 - 2(d-2)\mathbb{h}_4 - 2 \varphi \partial_\varphi \mathbb{h}_4\,,\label{eq:GGDef}\\
\mathbb{H}(\varphi,\chi) &=& \mathbb{h}_2\label{eq:HHDef}\,.
\ea
The remaining equation of motion~\eqref{eq:OmegaEOM} integrated to yield $\Omega(r,f) = 2M$  then amounts to the following algebraic equation determining $f$,
\ba\label{eq:EqMasterQTG}
r^{d-3} f \mathbb{G}(r,f) + \int \dd{r}r^{d-2}\,\mathbb{H}(r,f) &=& 2M\,.
\ea
This is our key result for generic quasi-topological gravities in $d\geq 4$.~\footnote{One may verify explicitly that this equation recovers the previously derived algebraic equations for $f$ in polynomial and non-polynomial CQTGs as a special subcase. Changing variables as in~\eqref{eq:VariableChange} and requiring $\mathbb{h}_i$ to depend only on $\psi$, the function $\mathbb{G}$ is given by
	\ba
	\mathbb{G}(\varphi,\psi) &=& \mathbb{h}_3(\psi) - 2 (d-2)\mathbb{h}_4(\psi) + 4 \psi \mathbb{h}_4'(\psi) \,\, = \,\, \mathbb{g}(\psi)\,.
	\ea
	Consider first $\mathbb{g} = 0$ and $\mathbb{h}_2$ given as in~\eqref{eq:hiPolynomial}. Then
	\ba 
\int \dd{r}	r^{d-2} \,\mathbb{h}_2(\psi(r,f)) &=& \int \dd{r} r^{d-2}\qty[(d-1)\mathbb{h} - 2 \psi \mathbb{h}' ] \nn\\
&=& \int \dd{r} r^{d-4} \qty[(d-1) r^2 \mathbb{h}\qty(\frac{k-f}{r^2}) + 2 (f-k) \mathbb{h}'\qty(\frac{f-k}{r^2})] \,\,=\,\, r^{d-1} \mathbb{h}\qty(\frac{k-f}{r^2})\,,\quad 
	\ea
	and we recover~\eqref{eq:EqMasterLocalPolynomialQTG} from~\eqref{eq:EqMasterQTG}. In the more general case when $\mathbb{g} \neq 0$, one may use the definitions~\eqref{eq:hDef}--\eqref{eq:gDef} 
	and the integrated version of~\eqref{eq:HiConstraint} to express $\mathbb{h}_2$ and analogously recover~\eqref{eq:EqMasterLocalNonPolynomialQTG}.
}
\\

We can collect the previous results in the following theorem. \\

{\bf{Theorem:}} {\it Let $\mathcal{L}\qty(g^{\mu\nu},R_{\mu\nu\rho\sigma},\nabla_\mu)$ be a $d$-dimensional gravitational theory 
	possessing second-order equations of motion  on~\eqref{eq:MetricToroidal}. Assume the two-dimensional Horndeski theory obtained from its reduction on these backgrounds satisfies the integrability condition~\eqref{eq:Integrability}, i.e., $\mathcal{L}$ is of quasi-topological type. Then the solutions satisfy
	\ba\label{eq:EOMTheorem2}
	\beta \partial_t f \,\,=\,\, 0\,,\,\,\,\quad \quad  \beta \partial_r n \,\,=\,\, 0\,,\,\,\,\quad \quad 
	\derivative{}{r}\qty[r^{d-3} f \mathbb{G}(r,f)  +\int {\dd{r}}\, r^{d-2}\,\mathbb{H}(r,f )]\,\,=\,\,0\,,
	\ea
where the functions $\mathbb{H}$ and $\mathbb{G}$ can be computed from the reduced Lagrangian density evaluated on a single-function static ansatz
\ba\label{eq:MetricStatick2}
g_{\mu\nu}(x)\dd{x}^\mu \dd{x}^\nu &=& - f(r)\dd{t}^2 + \frac{\dd{r}^2}{f(r)} + r^2 \dd{\Sigma_{d-2}^2}\,,
\ea
concretely} $\mathcal{L}_f \equiv\qty[ \varphi^{2-d} \mathcal{L}_{\text{\text{Horndeski}}}]_{\eqref{eq:MetricStatick2}}$ {\it defined through~\eqref{eq:SHorndeskiRewritten},
as follows,
	\ba
	\mathbb{H}\qty(r,f) &=& \eval{\mathcal{L}_f}_{f' = 0,\, f'' = 0} \,,\label{eq:HH}\\
	\mathbb{G}\qty(r,f) &=& - r\eval{\partial_{f'} \mathcal{L}_f}_{f' = 0}     + 2(d-2)\,\partial_{f''}\mathcal{L}_f+  2 \,r\,\partial_r \,\partial_{f''}\mathcal{L}_f\,.\label{eq:GG}
	\ea
}The last formula in~\eqref{eq:EOMTheorem2} is understood as first performing the $r$-integral in square brackets and subsequently replacing $f \to f(r)$ before taking the total derivative with respect to $r$.\\

 {\it Proof:} The first half of the proof is identical to the first half of the proof of the theorem stated in subsection~\ref{SecSub:NonPolynomialLocalQTG}. It remains to see that the characteristic function $\Omega(r,f)$ derived in~\eqref{eq:OmegaQuasiLocal} can be expressed onshell in terms of $\mathbb{H}$ and~$\mathbb{G}$ as in~\eqref{eq:HH}--\eqref{eq:GG}. These expressions may be verified by noticing that the reduced Lagrangian density evaluated on~\eqref{eq:MetricStatick2} is given by
\ba
\mathcal{L}_f &=& \mathbb{h}_2(r,f) - \mathbb{h}_3(r,f)\frac{f'}{r} - \mathbb{h}_4(r,f)f'' - 2 \, r^2 \,\partial_f \mathbb{h}_4(r,f) \frac{f'^2}{ r^2} \,.
\ea
Therefrom $\mathbb{H}$ and $\mathbb{G}$ defined in~\eqref{eq:HHDef}--\eqref{eq:GGDef} can be straightforwardly expressed in the form~\eqref{eq:HH}--\eqref{eq:GG}. 
\qed

\section{Reconstruction of static spherically symmetric solutions with $g_{tt}g_{rr} = -1$}\label{Sec:SingleFunctionStaticBHs}

The previous results imply 
 that any $d$-dimensional static spherically symmetric and asymptotically flat spacetime
\ba \label{eq:MetricStatickAgain}
g_{\mu\nu}(x)\dd{x}^\mu \dd{x^\nu} &=& -f(r)\dd{t}^2 + \frac{\dd{r}^2}{f(r)} + r^2 \dd{\Omega_{d-2}^2}
\ea
with an invertible dependence on the ADM mass can be generated as a solution to a purely gravitational $d$-dimensional theory. 
The ADM mass can be computed from the expansion of $f$ at asymptotic infinity as
\ba
f(r) &=& 1 - \frac{\mu}{r^{d-3}} + \cdots \,\,\, \quad \,\,\, \Rightarrow \,\,\,\quad \,\,\, M_{\text{ADM}} \,\, =\,\, \frac{(d-2) \Omega_{d-2} }{16 \pi G} \,\mu\,,
\ea
where $\Omega_{d-2}$ is the surface area of the $d-2$ dimensional unit sphere.
Inverting  $f = f(r;\mu)$ in the form $\mu=\mu(r,f)$, one may thus identify the parameter $\mu$ with the integration constant $2M$ arising in solutions of two-dimensional Horndesi theories satisfying the integrability condition~\eqref{eq:Integrability}, cf.~equation~\eqref{eq:OmegaEOM} which integrates to~\eqref{eq:OmegaOnshell}.~\footnote{Notice that, as a result, here and in what follows $M$ has dimension of $[\text{length}]^{d-3}$.} This then allows a posteriori the offshell identification of the characteristic function
\ba\label{eq:OmegaM}
\Omega(\varphi,\chi) = \mu(\varphi,\chi)\,.
\ea
Therefrom the generating functions $\alpha(\varphi,\chi)$ and $\beta(\varphi,\chi)$ which describe this configuration as a solution to the Horndeski equations of motion can be obtained from the defining relations~\eqref{eq:IntegrabilityOmega}. Such a reverse construction of solutions to two-dimensional Horndeski theories has been proposed and developed in~\cite{Boyanov:2025pes,Carballo-Rubio:2025ntd} and has also been applied in~\cite{Borissova:2026dlz,Borissova:2026wmn}.
A particular realisation of the functions $h_i(\varphi,\chi)$ in the corresponding Horndeski action was stated at the end of subsection~\ref{SecSub:SolvingIntegrableTheories} and can be found in~\cite{Borissova:2026dlz}.\\

We have shown here that the equations of motion on~\eqref{eq:MetricStatickAgain} of generic two-dimensional Horndeski theories
can be generated from the spherical reduction of $d$-dimensional gravitational actions by following the procedure described in subsection~\ref{SecSub:Expressing2DHorndeskiCovariant}. Thus it follows that in reverse-constructing a given configuration for $q_{ab}(y)$ and $\varphi(y)$ as in~\eqref{eq:MetricStatickAgain} as solution to a two-dimensional Horndeki theory, one is actually constructing it as a vacuum solution to a  $d$-dimensional gravity. In particular it is possible to explicitly state a $d$-dimensional gravitational action which yields a given such spacetime as a vacuum solution --- cf.~\eqref{eq:SQuasiLocal}. An exemplary application of this approach in the context of non-polynomial CQTGs in $d=4$ can be found in~\cite{Borissova:2026wmn}.

\subsection{Example: $d$-dimensional Hayward spacetime}\label{SecSub:Hayward}

The $d$-dimensional generalisation of the Hayward metric~\cite{Hayward:2005gi} is given by
\ba\label{eq:fHayward}
f(r) &=& 1- \frac{2 M r^2}{r^{d-1} + 2 M l^{2}}\,,
\ea
where $l$ is a regularisation length parameter. Depending on its value, this metric describes either a regular black hole or a horizonless compact object.~\footnote{This metric arises in notably in applications of renormalisation-group improvement in asymptotic safety~\cite{Bonanno:2000ep}, see also~\cite{Eichhorn:2022bgu,Platania:2023srt} for reviews.}\\

The Hayward metric has been obtained as a static spherically symmetric vacuum solution to polynomial CQTGs in odd dimensions $d\geq 5$~\cite{Bueno:2024dgm,Bueno:2024zsx} and to non-polynomial CQTGs in $d=4$~\cite{Borissova:2026wmn}. Here we review its characteristic function in order to distinguish this and other spacetimes which can be obtained from CQTGs, from the discussion of the  Bardeen spacetime later which cannot be obtained from CQTGs.\\

Inverting $f(r;M)$ in~\eqref{eq:fHayward} in the form $M(r,f)$ allows us to identify the characteristic  function $\Omega$ according to~\eqref{eq:OmegaM},
\ba
\Omega(\varphi,\psi) &=& 
 \frac{\varphi^{d-1}\psi}{1-l^{2} \psi}\,,
\ea
where we have applied the variable transformation~\eqref{eq:VariableChange}. See also~\cite{Boyanov:2025pes} and moreover~\cite{Carballo-Rubio:2025ntd,Borissova:2026dlz,Borissova:2026wmn} for the particular case of $d=4$. The dependence of the function $\Omega(\varphi,\psi)$ on $\varphi$ and $\psi$ is of the particular form~\eqref{eq:OmegaLocal}. This is a sufficient condition for the Hayward metric to arise as a solution to non-polynomial CQTGs as discussed for $d=4$ in~\cite{Borissova:2026wmn}. Moreover the algebraic equation~\eqref{eq:EqMasterLocalNonPolynomialQTG} is satisfied with
\ba\label{eq:hgHayward}
\mathbb{h}(\psi) \,\,=\,\, \frac{\psi}{1-l^{2} \psi}\,, \,\,\,\quad \quad \mathbb{g}(\psi) \,\,=\,\, 0\,.
\ea
We have seen that the last identity  is a necessary condition for a spacetime to arise as a solution to polynomial CQTGs. The sufficient condition for the Hayward spacetime to arise from such theories is that the function $\mathbb{h}$ can be represented as a resummation series of polynomial curvature quasi-topological densities in odd dimensions $d \geq 5$ evaluated onshell on~\eqref{eq:MetricStatickAgain}, in the form~\cite{Bueno:2024dgm}
\ba\label{eq:ResummationSeries}
\mathbb{h}(\psi) \,\,=\,\, \psi + \sum_{n=2}^{\infty} \alpha_n \psi^n\,,\quad \quad \,\,\, \alpha_n \,\,=\,\, l^{2(n-1)}\,,
\ea
which is a geometric series evaluating to~\eqref{eq:hgHayward}. It should be noted that in order to generate the Hayward spacetime as a solution to polynomial CQTGs, an infinite tower of curvature terms is needed, whereas in non-polynomial CQTGs this is not the case as one may generate functions $\mathbb{h}_i(\psi)$ in the reduced action which are arbitrary and not necessarily analytic, as observed in~\cite{Borissova:2026wmn}.

\subsection{Example: $d$-dimensional Dymnikova spacetime}\label{SecSub:Dymnikova}

The $d$-dimensional generalisation of the Dymnikova metric~\cite{Dymnikova:1992ux} is given by
\ba\label{eq:fDymnikova}
f(r) &=& 1-  \frac{2M}{r^{d-3}}\qty(1 - e^{-\frac{r^{d-1}}{2M l^{2} }})\,,
\ea
where $l$ is a regularisation length parameter. Depending on its value, the metric describes either a regular black hole or a horizonless compact object.~\footnote{This metric and variants thereof have also been obtained in the context of renormalisation-group improved black holes~\cite{Platania:2019kyx,Borissova:2022mgd}. We are here deriving these spacetimes as vacuum solutions to a $d$-dimensional generally covariant gravitational theory, the existence of which is a priori not clear when replacing the classical Newton coupling by a scale-dependent function at the level of the Schwarzschild spacetime.}\\

The Dymnikova metric has been discussed as a solution to non-polynomial CQTGs in $d=4$ in~\cite{Borissova:2026wmn}.\\

As observed in~\cite{Konoplya:2024kih}, an explicit inversion of $f(r;M)$ in~\eqref{eq:fDymnikova}  in the form $M(r,f)$ is possible in terms of the principal Lambert function $W_0$ satisfying the defining relations $W_0(z) = w \,:\Leftrightarrow \, z = w e^{w}$ and $W_0(0)=0$ as well as $W_0\qty(-e^{-1})=-1$. This allows us to identify the characteristic function $\Omega$ according to~\eqref{eq:OmegaM},
\ba
\Omega(\varphi,\psi) &=& \frac{\varphi^{d-1}\psi}{1 + l^{2} \psi \,W_0 \qty(- \frac{e^{-\frac{1}{l^{2} \psi}}}{l^{2} \psi})}\,.
\ea
The dependence of the function $\Omega(\varphi,\psi)$ on $\varphi$ and $\psi$ is of the particular form~\eqref{eq:OmegaLocal} and thus the Dymnikova spacetime arises explicitly as a solution to non-polynomial CQTGs. Even though the  algebraic equation~\eqref{eq:EqMasterLocalNonPolynomialQTG} can be satisfied with
\ba\label{eq:hgDymnikova}
\mathbb{h}(\psi) \,\,=\,\, \frac{\psi}{1 + l^{2} \psi \,W_0 \qty(- \frac{e^{-\frac{1}{l^{2} \psi}}}{l^{2} \psi})}\,, \,\,\,\quad \quad \mathbb{g}(\psi) \,\,=\,\, 0\,,
\ea
it is not possible to generate the Dymnikova spacetime from polynomial CQTGs as the function $\mathbb{h}$ cannot be represented as a resummation series of the form as in the first equation in~\eqref{eq:ResummationSeries}.

\subsection{Example: $d$-dimensional Bardeen spacetime}\label{SecSub:Bardeen}

The $d$-dimensional generalisation of the Bardeen metric~\cite{Bardeen:1968bh} is given by
\ba
f(r) &=& 1- \frac{2 M r^2}{\qty(r^{d-2} + g^{d-2})^{\frac{d-1}{d-2}}}\,,
\ea
where $g$ is a regularisation parameter which may be interpreted as a magnetic monopole charge in derivations of this spacetime as a solution to general relativity coupled to non-linear electrodynamics~\cite{Ayon-Beato:1998hmi,Ayon-Beato:1999kuh,Ayon-Beato:1999qin,Ayon-Beato:2000mjt}. Depending on its value, the metric describes either a regular black hole or a horizonless compact object. The Bardeen spacetime could not be obtained as a solution to polynomial CQTGs in $d \geq 5$~\cite{Bueno:2024dgm,Bueno:2025tli} and to  non-polynomial CQTGs in $d=4$~\cite{Borissova:2026wmn}, but can be obtained as a solution to the extended  QTGs in $d\geq 4$ discovered here.\\

To see this explicitly, we invert $f(r;M)$ in the form $M(r,f)$ and apply~\eqref{eq:OmegaM} to identify the characteristic function $\Omega$ offshell,
\ba
\Omega(\varphi,\chi) &=& \frac{1-\chi}{\varphi^2}\qty( \varphi^{d-2} + g^{d-2})^{\frac{d-1}{d-2}} \,\,=\,\, \psi \qty( \varphi^{d-2} + g^{d-2})^{\frac{d-1}{d-2}}\,,
\ea
where in the last step we have applied the variable transformation~\eqref{eq:VariableChange} to $\varphi$ and $\psi$. See also~\cite{Carballo-Rubio:2025ntd} for the particular case  of $d=4$. The dependence of the function $\Omega(\varphi,\psi)$ on $\varphi$ and $\psi$ is not of the particular form~\eqref{eq:OmegaLocal}, which explains 
why the Bardeen metric cannot be generated from CQTGs.~\footnote{A related observation concerns the failure of Markov's limiting curvature hypothesis~\cite{Markov:1982rcm} for the Bardeen spacetime.
	Even though curvature invariants are finite, they are not universally bounded for every solution and diverge as $M\to \infty$. In~\cite{Frolov:2024hhe} by a scaling argument it has been shown that if a regular black-hole spacetime satisfies the algebraic equation $\mathbb{h}(\psi) = 2M/r^{d-1}$, as does a vacuum solution to polynomial CQTGs, then this spacetime satisfies the limiting curvature conjecture. See also~\cite{Bueno:2024zsx,Bueno:2025tli} and moreover~\cite{Frolov:2021vbg} for a related sufficient condition. More recently~\cite{Bueno:2026dln} has performed a detailed analysis concerning the validity of the limiting curvature  conjecture in polynomial CQTGs coupled to matter. It will be interesting to generalise these discussions to the extended algebraic equation for $f$ arising in non-polynomial CQTGs~\eqref{eq:EqMasterLocalNonPolynomialQTG} and to the most general algebraic equation for $f$ arising in generic QTGs~\eqref{eq:EqMasterQTG}.} One may, however, find functions $\mathbb{G}(\varphi,\chi)$ and $\mathbb{H}(\varphi,\chi)$ which make this spacetime a solution to~\eqref{eq:EqMasterQTG}. An exemplary choice is
\ba
\mathbb{H}(\varphi,\chi) \,\,=\,\, \frac{1-\chi}{\varphi^{d+1}}  \qty(\varphi^{d-2}+g^{d-2})^{\frac{d-1}{d-2}}\, \frac{(d-3) g^2 \varphi^d - 2 g^d \varphi^2}{ g^2 \varphi^d + g^d \varphi^2} \,, \,\,\, \quad \quad \mathbb{G}(\varphi,\chi) \,\,=\,\, 0\,.
\ea
This establishes the Bardeen spacetime as a solution to  extended QTGs involving in general curvature and curvature-derivative invariants.

\section{Discussion}\label{Sec:Discussion}

We have shown that the generic two-dimensional Horndeski theories relevant for the description of $d$-dimensional black holes can arise from the reduction on $2+(d+2)$ warped-product backgrounds of purely gravitational theories in $d\geq 4$ dimensions. This makes the corresponding onshell configurations for the two-dimensional metric and scalar field genuine gravitational vacuum solutions. Our main motivation stems from the possibility of thereby reconstructing $d$-dimensional  static spherically symmetric and asymptotically flat regular black-hole spacetimes satisfying $g_{tt}g_{rr} = -1$ as vacuum solutions of a $d$-dimensional gravitational theory. This applies to any such spacetime with an invertible dependence on the ADM mass regardless of the specific form of the inversion --- provided the gravitational action is allowed to depend non-polynomially on curvature and curvature-derivative invariants. The solution space of these here discovered extended $d\geq 4$ dimensional QTGs is profoundly greater than the one of polynomial CQTGs which exist only in $d\geq 5$ dimensions, see~e.g.~the solutions and applications in~\cite{Oliva:2010eb,Myers:2010jv,Dehghani:2011vu,Cisterna:2017umf,Bueno:2019ltp,Bueno:2019ycr, Bueno:2022res,Moreno:2023rfl,Bueno:2024dgm,DiFilippo:2024mwm,Frolov:2024hhe,Bueno:2024zsx,Bueno:2024eig,Bueno:2025gjg,Aguayo:2025xfi,Hennigar:2025yqm,Bueno:2025qjk,Bueno:2025tli,Hao:2025utc,Frolov:2026rcm,Tsuda:2026xjc,Li:2026mam,PinedoSoto:2026hfm}, as well as the one of $d=4$ non-polynomial CQTGs discussed  in, see~e.g.~\cite{Borissova:2026wmn,Bueno:2025zaj,Konoplya:2026gim,Lutfuoglu:2026gis,Dubinsky:2026wcv,Malik:2026laq}. In particular the metric function $f$ of static spherically symmetric solutions in this case is determined by an algebraic equation involving an arbitrary functional dependence on $f$ and $r$. Thereby regular black holes, which would generically require the coupling for instance to non-linear electrodynamic matter sources~\cite{Ayon-Beato:1998hmi,Ayon-Beato:1999qin,Ayon-Beato:1999kuh,Ayon-Beato:2000mjt,Bronnikov:2000vy,Dymnikova:2004zc,Bronnikov:2022ofk,Balart:2014cga,Fan:2016hvf,Rodrigues:2018bdc}, can be generated as vacuum solutions. One may therefore in an abstract sense view the gravitational theories proposed here as effective field theories obtained by integrating out such matter fields from a gravity-matter theory.\\

As a key application one may consider  coupling two-dimensional external matter sources as the spherical reduction of $d$-dimensional ones and employ the geometrodynamic framework of two-dimensional Horndeski theory to describe $d$-dimensional non-singular gravitational collapse. 
This is perhaps one of the primary motivations for the generalised effective Einstein equations on the space of $2+(d-2)$ warped-product backgrounds constructed in~\cite{Carballo-Rubio:2025ntd},
\ba\label{eq:EffectiveEinstein}
\mathbb{G}_{\mu\nu}(q_{ab},\varphi) &=& 8 \pi G T_{\mu\nu}\,,
\ea
where $\mathbb{G}$ is a second-order symmetric and covariantly conserved tensor defined from the variations of the two-dimensional Horndeski action as
\ba\label{eq:GGEinstein}
\mathbb{G}_{\mu\nu}(q_{ab},\varphi) &=& \varphi^{2-d} \mathcal{E}_{ab} \delta^a_\mu \delta^b_\nu - \frac{1}{2(d-2)}\varphi^{5-d} \mathcal{E}_\varphi \gamma_{ij} \delta^i_\mu \delta^j_\nu \,,
\ea
and $T$ is a $d$-dimensional external source. Regular black-hole formation has been studied in the context of polynomial CQTGs in $d\geq 5$ dimensions~\cite{Bueno:2024zsx,Bueno:2024eig,Bueno:2025gjg} and in the context of non-polynomial CQTGs in $d=4$ dimensions~\cite{Bueno:2025zaj}, whereby in all cases the dynamical equations are included as a subset in the set of equations~\eqref{eq:EffectiveEinstein}--\eqref{eq:GGEinstein}.
Further applications of regular black-hole formation described by the above generalised effective Einstein equations have been considered in~\cite{Boyanov:2025pes,Borissova:2026dlz}. The main statement which follows from our discussion here is that solving~\eqref{eq:EffectiveEinstein} for generic $q_{ab}$ and $\varphi$ amounts to actually solving a $d$-dimensional generally covariant  gravitational theory coupled to matter. In particular, for each choice of two-dimensional metric $q_{ab}$ and scalar field $\varphi$ one may find explicitly a $d$-dimensional gravitational action with second-order equations of motion on $2+(d-2)$ warped-product backgrounds, whose equation of motion tensor $\mathcal{E}_{\mu\nu}$ obtained by variation with respect to the spacetime metric $g_{\mu\nu}$ and evaluated on a $2+(d-2)$ warped-product ansatz is proportional to~\eqref{eq:GGEinstein}, i.e., $\eval{\mathcal{E}_{\mu\nu}}_{\eqref{eq:Metric}} \propto \mathbb{G}_{\mu\nu}$. We thus view~\cite{Carballo-Rubio:2025ntd} in combination with this work as a compelling pathway towards addressing central challenges and open questions concerning black holes in different theories and dimensions~\cite{Carballo-Rubio:2025fnc,Bambi:2025wjx,Buoninfante:2024oxl}. Particularly appealing future directions are a unified framework for the thermodynamics of $d$-dimensional static black holes satisfying $g_{tt}g_{rr}=-1$ in Schwarzschild gauge based on Wald's Noether charge formalism~\cite{Wald:1993nt,Iyer:1994ys,Jacobson:1993vj}, as initiated in~\cite{Borissova:2026rbi}, or based on the Euclidean path integral approach~\cite{Gibbons:1976ue,Hawking:1978jz}, as well as an advanced understanding of the impact of  singularity theorems~\cite{Penrose:1964wq,Hawking:1970zqf,Hawking:1973uf} beyond general relativity. \\

Another application concerns dilaton models discussed in string theory and holographic approaches such as~\cite{Mandal:1991tz,Witten:1991yr,Dijkgraaf:1991ba,McGuigan:1991qp,Ishibashi:1991wh,DeAlwis:1991vz,Khastgir:1991ip,Callan:1992rs}. It will be interesting to consider  the reconstruction of gravitational effective actions resulting in these models or the deduction of universal structures that such effective actions must possess. One may expect such discussions to provide a novel way of addressing swampland criteria~\cite{Vafa:2005ui,Ooguri:2006in,Palti:2019pca}.

\begin{acknowledgments}
\noindent I am deeply grateful to Ra{\'u}l Carballo-Rubio for invaluable discussions. I also wish to thank Aimeric Coll\'eaux for very useful written communication.
This work is supported by STFC Consolidated Grant ST/X000575/1.
\end{acknowledgments}

\bibliographystyle{jhep}
\bibliography{references}

\end{document}